\documentclass[conference]{IEEEtran}
\IEEEoverridecommandlockouts
\usepackage{cite}
\usepackage{color}
\usepackage{amsmath,amssymb,amsfonts}
\usepackage{algorithmic}
\usepackage{graphicx}
\usepackage{textcomp}
\usepackage{subfig}

\usepackage{xcolor}
\usepackage{colortbl}

\def\BibTeX{{\rm B\kern-.05em{\sc i\kern-.025em b}\kern-.08em
    T\kern-.1667em\lower.7ex\hbox{E}\kern-.125emX}}

\usepackage[ruled]{algorithm2e}

\usepackage{url}

\def\BibTeX{{\rm B\kern-.05em{\sc i\kern-.025em b}\kern-.08em T\kern-.1667em\lower.7ex\hbox{E}\kern-.125emX}}

\newcommand {\mymarginpar}[1]{\marginpar{#1}}
\renewcommand {\marginpar}[1]{}

\def\_{\rule{.3em}{.15ex}}      

\newcommand{\ls}[1]
   {\dimen0=\fontdimen6\the\font
    \lineskip=#1\dimen0
    \advance\lineskip.5\fontdimen5\the\font
    \advance\lineskip-\dimen0
    \lineskiplimit=.9\lineskip
    \baselineskip=\lineskip
    \advance\baselineskip\dimen0
    \normallineskip\lineskip
    \normallineskiplimit\lineskiplimit
    \normalbaselineskip\baselineskip
    \ignorespaces
   }

\newcommand {\bearn}{\begin{eqnarray*}}
\newcommand {\eearn}{\end{eqnarray*}}
\newcommand {\barr}{\begin{array}}
\newcommand {\earr}{\end{array}}

\newcommand {\N}{{\cal N}}

\newtheorem{definition}{Definition}
\newtheorem{property}[definition]{Property}
\newtheorem{proposition}[definition]{Proposition}
\newtheorem{lemma}[definition]{Lemma}
\newtheorem{theorem}[definition]{Theorem}
\newtheorem{corollary}[definition]{Corollary}
\newtheorem{example}{Example}
\newtheorem{remark}[definition]{Remark}

\newcommand {\benum} {\begin{enumerate}}
\newcommand {\eenum} {\end{enumerate}}

\newcommand {\bdesc} {\begin{description}}
\newcommand {\edesc} {\end{description}}

\newcommand {\bfig}[2] {\begin{figure}
  \centering
  \includegraphics[width=#2]{#1}}
\newcommand {\brotatefig}[2] {\begin{figure}[htbp]
                        \centerline {
                         \epsfig{figure={#1},clip=,angle=-90,width={#2}}}}
\newcommand {\bfigfirst}[2] {\begin{figure}[h]
                        \centerline {
                        \setlength{\epsfxsize}{#2}
                        \epsffile{#1}}}
\newcommand {\efig}[2]{ \caption{#2}
                        \label{fig:#1}
                        \end{figure}
                        \mymarginpar{fig:#1}}
\newcommand {\erotatefig}[2]{ \caption{#2}
                        \label{fig:#1}
                        \end{figure}
                        \mymarginpar{fig:#1}}
\newcommand {\rfig}[1]{Figure \ref{fig:#1}}

\newcommand {\btab}[1]{
                       \begin{table}
                       \centering
                       \begin{tabular}{#1}}
\newcommand {\etab}[3] {
                       \end{tabular}
                       \caption[#3]{#2}
                       \label{tab:#1}
                       \end{table}
                       \mymarginpar{tab:#1}
                       \vspace{.1in}}
\newcommand {\rtab}[1]{Table \ref{tab:#1}}

\newcommand {\btabular}[1]{\begin{center}
                       \begin{tabular}{#1}}
\newcommand {\etabular}{\end{tabular}
                       \end{center}}

\newcommand {\bdefin}[1]{\begin{definition}
                      \mymarginpar{def:#1}
                      \label{def:#1} }
\newcommand {\edefin}       {\end{definition}}

\newcommand {\bpro}[1]{\begin{property}
                      \mymarginpar{pro:#1}
                      \label{pro:#1} }
\newcommand {\epro}   {\end{property}}

\newcommand {\bprop}[1]{\begin{proposition}
                      \mymarginpar{prop:#1}
                      \label{prop:#1} }
\newcommand {\eprop}       {\end{proposition}}

\newcommand {\blem}[1]{\begin{lemma}
                      \mymarginpar{lem:#1}
                      \label{lem:#1} }
\newcommand {\elem}   {\end{lemma}}
\newcommand {\rlem}[1]{Lemma \ref{lem:#1}}

\newcommand {\bthe}[1]{\begin{theorem}
                      \mymarginpar{the:#1}
                      \label{the:#1} }
\newcommand {\ethe}   {\end{theorem}}
\newcommand {\rthe}[1]{Theorem \ref{the:#1}}

\newcommand {\bproof}{\noindent {\bf Proof.} \ }
\newcommand {\eproof} {\hfill \squares \\ \vspace{.3cm}}
\newcommand {\bcor}[1]{\begin{corollary}
                      \mymarginpar{cor:#1}
                      \label{cor:#1} }
\newcommand {\ecor}   {\end{corollary}}

\newcommand {\bax}[1]{\begin{axiom}
                      \mymarginpar{ax:#1}
                      \label{ax:#1} }
\newcommand {\eax}       {\vspace{-.1in} \end{axiom}}

\newcommand {\bex}[2]{\vspace{.1in}
                      \begin{example}
                      \mymarginpar{ex:#1}
                       {\bf #2}
                      \label{ex:#1} }
\newcommand {\eex}       {\end{example} \vspace{.3cm} }
\newcommand {\rex}[1]{Example \ref{ex:#1}}

\newcommand {\brem}[1]{\begin{remark}
                      \mymarginpar{rem:#1}
                      \label{rem:#1} \em }
\newcommand {\erem}   {\end{remark}}

\newcommand {\beq}[1]{\mymarginpar{eq:#1}
                      \begin{equation}
                      \label{eq:#1} }

\newcommand {\beqno}[1]{\mymarginpar{eq:#1}
                      \begin{eqnarray}
                      \nonumber}

\newcommand {\eeq}       {\end{equation}}
\newcommand {\eeqno}       { && \end{eqnarray}}
\newcommand {\req}[1]{(\ref{eq:#1})}

\newcommand {\bear}[1]{\mymarginpar{eq:#1}
                       \begin{eqnarray}
                       \label{eq:#1} }

\newcommand {\bearno}[1]{\mymarginpar{eq:#1}
                       \begin{eqnarray}
                       \nonumber}

\newcommand {\eear}{\end{eqnarray}}
\newcommand {\eearno}{\end{eqnarray}}
\newcommand {\bsel}{\left \{ \begin{array}{cl}}
\newcommand {\esel}{\end{array} \right.}

\newcommand {\bmat}[1]{\left [ \begin{array}{#1}}
\newcommand {\emat}{\end{array} \right ]}
\newcommand {\bsec}[2]{\mymarginpar{sec:#2}
                       \section{#1}
                       \label{sec:#2} }

\newcommand {\rsec}[1]{Section \ref{sec:#1}}

\newcommand {\bsubsec}[2]{\mymarginpar{sec:#2}
                       \subsection{#1}
                       \label{sec:#2} }

\newcommand {\rsubsec}[1]{Section \ref{sec:#1}}
\newcommand{\bsubsubsec}[2]{\mymarginpar{subsubsec:#2}\subsubsection{#1}\label{subsubsec:#2}}
\newcommand{\rsubsubsec}[1]{Section \ref{subsubsec:#1}}

\def\R{I\kern-0.30em R}
\def\N{I\kern-0.30em N}
\def\P{I\kern-0.30em P}
\newcommand\squares{\vrule height6pt width7pt depth1pt}

\def\ex{{\bf\sf E}}
\def\pr{{\bf\sf P}}

\newcommand{\betak}{{r_{0,k}}}

\newcommand\sgo{(G(V,E), p_{U,W}(\cdot,\cdot))}

\newcommand\qq{q}
\newcommand \Q {{\bf Q}}

\begin{document}

\title{Positively Correlated Samples Save Pooled Testing Costs}

\author{Yi-Jheng~Lin,
		Che-Hao~Yu,
		Tzu-Hsuan~Liu,
		Cheng-Shang~Chang,~\IEEEmembership{Fellow,~IEEE,}\\
		and Wen-Tsuen~Chen,~\IEEEmembership{Life Fellow,~IEEE}
        \IEEEcompsocitemizethanks{\IEEEcompsocthanksitem
        The authors are with the Institute of Communications Engineering, National Tsing Hua University, Hsinchu 30013, Taiwan R.O.C.

        Email:  s107064901@m107.nthu.edu.tw; chehaoyu@gapp.nthu.edu.tw;  carina000314@gmail.com;  cschang@ee.nthu.edu.tw;  wtchen@cs.nthu.edu.tw.
    	}}

\maketitle
\begin{abstract}
	The group testing approach that achieves significant cost reduction over the individual testing approach has received a lot of interest lately for massive testing of COVID-19. Many studies simply assume samples mixed in a group are independent. However, this assumption may not be reasonable for a contagious disease like COVID-19. Specifically, people within a family tend to infect each other and thus are likely to be positively correlated. By exploiting positive correlation, we make the following two main contributions. One is to provide a rigorous proof that further cost reduction can be achieved by using the Dorfman two-stage method when samples within a group are positively correlated. The other is to propose a hierarchical agglomerative algorithm for pooled testing with a social graph, where an edge in the social graph connects frequent social contacts between two persons. Such an algorithm leads to notable cost reduction (roughly 20\%-35\%) compared to random pooling when the Dorfman two-stage algorithm is applied.

\end{abstract}

{\bf Keywords:} COVID-19, group testing, regenerative processes, Markov modulated processes, social networks.




%

\bsec{Introduction}{introduction}

Massive testing is one of the most effective measures to detect and isolate asymptomatic COVID-19 infections so as to
reduce the transmission rate of COVID-19 \cite{chen2020time}.
However, massive testing for a large population is very costly if it is done one at a time.
The recent article posted on the US FDA website \cite{fdaAug} indicates that the group testing approach (or pool testing, pooled testing, batch testing) has received a lot of interest lately. Such an approach
(testing a group of mixed samples) can greatly save testing resources for a population with a low prevalence rate \cite{lohse2020pooling,abdalhamid2020assessment,yelin2020evaluation,gollier2020group}.
Moreover,  the following  testing procedure is suggested in the US CDC's guidance for the use of pooling procedures in SARS-CoV-2 \cite{cdcAug}:

{\em
``If a pooled test result is negative, then all specimens can be presumed negative with the single test. If the test result is positive or indeterminate, then all the specimens in the pool need to be retested individually.''}

A simple testing procedure that implements the above guidance is known as  Dorfman's two-stage group testing method \cite{dorfman1943detection}. The method first partitions the population into groups of $M$ samples. If the test of a group of $M$ samples is negative, then all the $M$ samples in that group are declared to be negative. Otherwise, each sample in that group is retested individually.
Such a method has been implemented by many countries for massive testing of COVID-19 \cite{wiki2020list}.

To measure the amount of saving of a group testing method, Dorfman used the {\em expected relative cost} (that is defined as the ratio of the expected number of tests required by the group testing method to the number of tests required by the individual testing).
The expected relative cost for
independent and identically distributed (i.i.d.) samples was derived in \cite{dorfman1943detection}.
Suppose that the prevalence rate (the probability that a randomly selected sample is positive) is $r_1$.
Note that if the test result of a group is positive, all the samples in that group need to be retested individually.
For a group of $M$ samples, the group is tested positive with the probability $1-(r_0)^M$, where $r_0=1-r_1$.
So the expected number of tests for the group is $1+M \cdot (1-(r_0)^M) = (M+1)-M(r_0)^M$.
Thus, the expected relative cost for i.i.d. samples with group size $M$ is
	\beq{dorf1111}
	\frac{M+1}{M}-(r_0)^M.
	\eeq
One can then use \req{dorf1111} to optimize the group size $M$ according to the prevalence rate \cite{dorfman1943detection}.

There are more sophisticated group testing methods for implementing the CDC's guidance for testing COVID-19 (see e.g.,
\cite{sinnott2020evaluation,shental2020efficient,ghosh2020tapestry,ghosh2020compressed,mutesa2020pooled,lin2020comparisons}). These methods require diluting a sample and then pooling the diluted samples into multiple groups (pooled samples). Such methods are specified by two components:
(i) a {\em pooling matrix} that directs each diluted sample to be pooled into a specific group, and (ii) a {\em decoding algorithm} that uses the test results of pooled samples to reconstruct the status (i.e., a positive or negative result) of each sample. As shown in the recent comparative study \cite{lin2020comparisons}, the expected relative costs of such methods depend heavily on the pooling matrix, and one has to select an appropriate pooling matrix according to the prevalence rate. For i.i.d. samples, using such sophisticated methods result in significant gains over the simple Dorfman two-stage group testing method, in particular when the prevalence rate is low (below 5\%).

In practice, samples are not i.i.d. For a contagious disease like COVID-19, people in the same family (or social bubble) are likely to infect each other. 
Lendle {\it et al.} \cite{lendle2012group} studied the efficiency (i.e., the expected relative costs) for group testing methods when samples within a group are positively correlated exchangeable random variables. They derived closed-form expressions of efficiency for hierarchical- and matrix-based group testing methods under certain assumptions, and examined three models of exchangeable binary random variables. They concluded that positive correlations between samples within a group could improve efficiency.  
Moreover, in the recent WHO research article \cite{deckert2020simulation}, it was shown by computer simulations that pooled samples from {\em homogeneous groups of similar people} could lead to cost reduction for the Dorfman two-stage method.
The main objective of this paper is to provide insight and proof for that observation through a mathematical model.

Let us consider a testing site where people form a line (or queue) to be tested. 
It is reasonable to assume that people arriving in groups of various sizes are in contiguous positions of the line.	Since the disease prevalence rate in two arriving groups may differ, we say that two groups are of the same type if they have the same prevalence rate.
People in $M$ contiguous positions are pooled together and tested by using Dorfman's two-stage group testing method. For our analysis, we make the following three mathematical assumptions:
\begin{description}
	\item[(A1)] i.i.d. group sizes:
	The sizes of arriving groups of people are i.i.d. with a finite mean.
	\item[(A2)] i.i.d. group types:
	There are $K$ types of arriving groups. The types of arriving groups of people are i.i.d.
	With probability $\pi_k$, a group of arriving people is of type  $k$, $k=1,2, \ldots, K$.
	\item[(A3)] Homogeneous samples within the same group:
	Samples obtained from people within the same group are i.i.d. Bernoulli random variables with the same prevalence rate. With probability $\betak$ (resp. $r_{1,k}=1-\betak$),
	a sample in a type $k$ group is negative (resp. positive).
\end{description}

\begin{figure}
	\centering
	\includegraphics[width=0.45\textwidth]{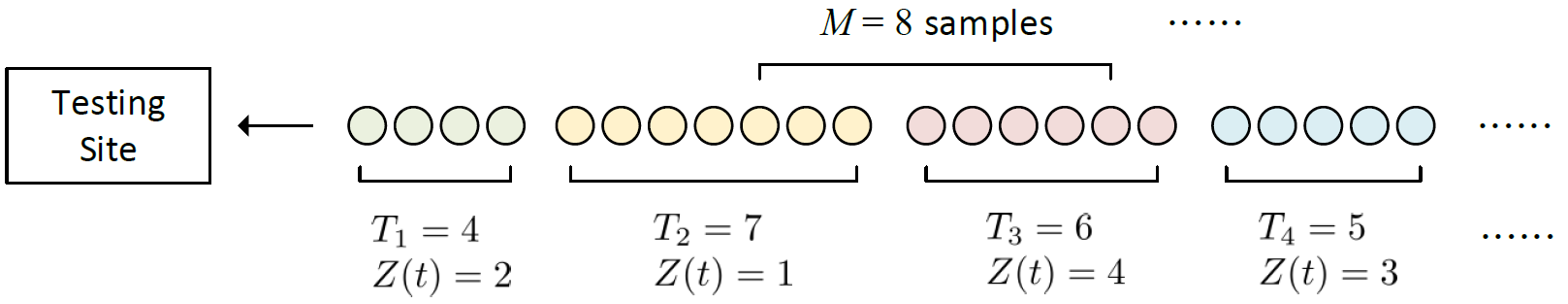}
	\caption{An illustration of an arrival process in a testing site, where $T_i$ is the group size of the $i^{th}$ arriving groups, $Z(t)$ is the group type of the $t^{th}$ sample, and $M$ samples of contiguous positions are pooled together for Dorfman's two-stage group testing.}
	\label{fig:queue}
\end{figure}

An illustration of an arrival process in a testing site is provided in \rfig{queue}.
In this figure, the number of people in the first group $T_1$ is $4$, the number of people in the second group $T_2$ is $7$, the number of people in the third group $T_3$ is $6$, and the number of people in the forth group $T_4$ is $5$. Eight samples of contiguous positions are pooled together for Dorfman's two-stage group testing, i.e., $M=8$.

Denote by $X(t)$ the indicator random variable of the $t^{th}$ sample in the line of the testing site. We say the $t^{th}$ sample is negative (resp. positive) if $X(t)=0$ (resp. $X(t)=1$).
Consider using the Dorfman two-stage method for testing the $M$ consecutive samples $X(t+1), X(t+2), \ldots, X(t+M)$  for some fixed $t \ge 0$.
With probability
$$1-\pr(X(t+1)=0,X(t+2)=0, \ldots, X(t+M)=0),$$
the test result for the group of $M$ consecutive samples
is positive and they need to be tested individually. Thus, the expected number of tests is
$$1+(1-\pr(X(t+1)=0,X(t+2)=0, \ldots, X(t+M)=0))M.$$
As such,
the expected relative cost for these $M$ samples by the Dorfman two-stage method is
\beq{dorf4444}
\frac{M+1}{M}- \pr(X(t+1)=0,X(t+2)=0, \ldots, X(t+M)=0).
\eeq

We state  the first main result of this paper in the following theorem.

\bthe{mma}
Suppose that the arriving process $\{X(t), t \ge 1\}$ satisfying (A1)-(A3).
The expected relative cost for pooling any $M$ consecutive samples into a group is not higher than
that for pooling $M$ samples at random, i.e.,
the expected relative cost in \req{dorf4444} is not higher than \req{dorf1111} with
\beq{dorf3333}
r_0=\sum_{k=1}^K \pi_k \betak.
\eeq
\ethe

Our second main result is the monotonicity of the expected relative cost under a stronger assumption than (A1).

\begin{description}
	\item[(A$1^+$)]
	The group sizes are independent and geometrically distributed with parameter $1-\omega$ for some $0 \le \omega \le 1$.
\end{description}

\bthe{geo} Suppose that the arriving process $\{X(t), t \ge 1\}$ satisfying (A$1^+$), (A2), and (A3).
Then the expected relative cost in \req{dorf4444} is decreasing in $\omega$.
\ethe

Note that when $\omega=0$, $\{X(t), t \ge 1\}$ is reduced to the sequence of i.i.d. samples with the prevalence rate $r_1$.
As such, the monotonicity result in \rthe{geo} is a stronger result than that in \rthe{mma}.

Our third main result is a closed-form expression for the expected relative cost under (A$1^+$), (A2), and (A3).

\bthe{mma2}
Under (A$1^+$), (A2), and (A3),
the expected relative cost is
\beq{trans88880}
\frac{M+1}{M}-{\bf \pi} {\bf R}  { ( {\bf P}{\bf R} )^{M-1} } {\bf 1},
\eeq
where
${\bf P}=(p_{i,j})$ is the $K \times K$  matrix with
\beq{trans11110}
p_{i,j}=\left\{\begin{array}{ll}
	\omega+(1-\omega) \pi_i &\mbox{if $j=i$}\\
	(1-\omega) \pi_j &\mbox{if $j\ne i$}
\end{array}
\right.,
\eeq
${\bf R}$ is the diagonal matrix with
the $k^{th}$ diagonal element being $r_{0,k}$,
${\bf 1}$ is the $K \times 1$ (column) vector with all its elements being 1,
and ${\bf \pi}$ is the $1 \times K$ (row) vector with its $k^{th}$ element being $\pi_k$.
\ethe

We can further derive the lower bound of the expected relative cost in \req{trans88880}.
	\bthe{mma4}
	Under (A$1^+$), (A2), and (A3), the expected relative cost is lower bounded by 
	\beq{lower1234}
	\frac{M+1}{M}-r_0 (\omega+(1-\omega) r_0)^{M-1}.
	\eeq
	\ethe

Using the closed-form expression in \rthe{mma2}, we compare the expected relative cost of the simple Dorfman two-stage method with the lowest expected relative cost of the $(d_1,d_2)$-regular pooling matrix \cite{lin2020comparisons}.
With a moderate positive correlation, our numerical results demonstrate that the gain by such a simple method outperforms those by using sophisticated strategies with $(d_1,d_2)$-regular pooling matrices when the prevalence rate is higher than $5\%$.

The results for samples in a line of a testing site only exploits the positive correlations between two contiguous samples in a line graph.
One important extension is to consider pooled testing with a social graph, where frequent social contacts between two persons are connected by an edge in the social graph. Contagious diseases such as COVID-19 can propagate the disease from an infected person to another person through the social contacts between two persons, two persons connected by an edge are likely to infect each other, and they are likely to be positively correlated. To exploit the positive correlation in a social graph, we adopt the probabilistic framework of sampled graphs for structural analysis in \cite{chang2011general,chang2015relative,chang2017probabilistic}. In particular, we propose a hierarchical agglomerative algorithm for pooled testing with a social graph (see Algorithm \ref{alg:hierarchicalgroup}). Our numerical results show that such an algorithm leads to significant cost reduction (roughly 20\%-35\%) compared to random pooling when the Dorfman two-stage algorithm is used.

The paper is organized as follows:  in \rsubsec{reg}, we prove \rthe{mma} and \rthe{geo} by using the renewal property of regenerative processes.
We then prove \rthe{mma2} and \rthe{mma4} in \rsubsec{mma} by using the Markov property of Markov modulated processes.
In \rsec{graph}, we extend the dependency of samples from a line graph to a general graph. There we propose a hierarchical agglomerative algorithm to exploit the positive correlation of samples.
The numerical results are shown in \rsec{num}.  The paper is concluded in \rsec{con}, where we discuss possible extensions for future works.

\bsec{Mathematical Analyses and Proofs}{proofs}

\bsubsec{Regenerative processes}{reg}

In this section, we prove the main result in \rthe{mma} and \rthe{geo} by using the renewal property of regenerative processes (see, e.g., Section 6.3 of the book \cite{nelson2013probability}).

Let $\{T_i, i \ge 1\}$ be the number of samples in the $i^{th}$ group, and $\tau_i=\sum_{\ell=1}^i T_\ell$ be the cumulative number of samples in the first $i$ groups.
Since we assume that $\{T_i, i \ge 1\}$ are i.i.d. in (A1), $\{\tau_i+1, i \ge 1\}$ is a renewal process. From (A2) and (A3), $\{X(t), t \ge 1\}$ is a regenerative process with the regenerative points $\{\tau_i+1, i \ge 1\}$, i.e., $\{X(\tau_i+t), t \ge 1\}$ has the same joint distribution as
$\{X(t), t \ge 1\}$.

In the following lemma, we derive the prevalence rate.

\blem{prev}
The prevalence rate of a randomly selected sample for the arrival process satisfying (A1)-(A3) is
\beq{dorf2222}
r_1= \sum_{k=1}^K \pi_k (1-\betak).
\eeq
Thus, $r_0=1-r_1=\sum_{k=1}^K \pi_k \betak$.
\elem

\bproof
Let $Z(t)$ be the group type of the $t^{th}$ sample. In view of (A2),
we have
\beq{dorf1122}
\pr(Z(t)=k)=\pi_k,
\eeq
Also, from (A3),
\beq{dorf1133}
\pr(X(t)=1 |Z(t)=k)=(1-\betak).
\eeq
From  the law of total probability, it follows that
\bear{dorf2222a}
\pr (X(t)=1)&=&\sum_{k=1}^K \pr(X(t)=1 |Z(t)=k) \pr(Z(t)=k)
\nonumber \\
&=&\sum_{k=1}^K \pi_k (1-\betak).
\eear
As \req{dorf2222a} holds for any arbitrary $t$, the prevalence rate of a randomly selected sample
is the same as \req{dorf2222a}.
\eproof

Now we prove \rthe{mma}.

\bproof (\rthe{mma})
In view of \req{dorf4444},
it suffices to show that for any $t \ge 0$,
\beq{mma9911}
\pr (X(t+1)=0, X(t+2)=0, \ldots , X(t+M)=0) \ge ( r_0)^{M}.
\eeq

For this, we first show that \req{mma9911} holds for $t=0$
by induction on $M$. Since $\pr(X(1)=0)=r_0$ from \rlem{prev},
the inequality in \req{mma9911} holds trivially for $M=1$.
Assume that the inequality in
\req{mma9911} holds for $t=0$ and all $m \le M-1$ as the induction hypothesis.
From the law of total probability, we have
\bear{mma9944}
&&\pr (X(1)=0,  \ldots , X(M)=0) \nonumber\\
&=&\sum_{s=1}^\infty \pr (X(1)=0,  \ldots , X(M)=0 |T_1=s)\pr (T_1=s) \nonumber\\
&=&\sum_{s=1}^{M-1} \pr (X(1)=0,  \ldots , X(M)=0 |T_1=s)\pr (T_1=s) \nonumber\\
&+&\sum_{s=M}^\infty \pr (X(1)=0,  \ldots , X(M)=0 |T_1=s)\pr (T_1=s).\nonumber\\
\eear
Conditioning on the event $\{T_1=s\}$ for $s \ge M$, the number of samples in the first group is not smaller  than $M$.
Thus, for $s\ge M$, we have from (A2) and (A3) that
\bear{mma9955}
&&\pr (X(1)=0,  \ldots , X(M)=0 |T_1=s)\nonumber\\
&&=\sum_{k=1}^K \pi_k (\betak)^M  \nonumber\\
&&\ge (\sum_{k=1}^K \pi_k \betak )^M=r_0^M,
\eear
where the last inequality follows from Jensen's inequality for the convex function $x^M$.
For $T_1=s \le M-1$, we know that the second group starts from $s+1$.
It then follows from the renewal property in (A1) that
\bear{mma9966}
&&\pr (X(1)=0,  \ldots , X(M)=0 |T_1=s)\nonumber\\
&&=\pr (X(1)=0,  \ldots , X(s)=0 |T_1=s)\nonumber\\
&&\quad\pr (X(s+1)=0,  \ldots , X(M)=0 |T_1=s)\nonumber\\
&&=\Big (\sum_{k=1}^K \pi_k (\betak)^s \Big ) \pr (X(1)=0,  \ldots , X(M-s)=0 ) \nonumber\\
&&\ge \Big(\sum_{k=1}^K \pi_k \betak\Big)^{s}\pr (X(1)=0,  \ldots , X(M-s)=0 ) \nonumber\\
&&\ge  (r_0)^{s} (r_0)^{M-s}=(r_0)^M
\eear
where the second last inequality follows from Jensen's inequality for the convex function $x^s$,
and the last inequality follows from
the induction hypothesis.
Using \req{mma9955} and \req{mma9966} in \req{mma9944} completes the induction for $t=0$ in \req{mma9911}.

Now we show that \req{mma9911} hold for any arbitrary $t$. For a fixed $t$, let $\tilde T_1(t)$ be the residual life from $t$ to the next regenerative point, i.e., the number of remaining samples in the same group of the $t^{th}$ sample.
The argument for any arbitrary $t$ then follows from the same inductive proof for $t=0$ by replacing $T_1$ with $\tilde T_1(t)$.
\eproof

In the proof of \rthe{mma}, we show that
\bear{mma9911b}
&&\pr (X(t+1)=0, X(t+2)=0, \ldots , X(t+M)=0) \nonumber\\
&&\ge ( r_0)^{M}=(\pr(X(1)=0))^M.
\eear
By replacing $\betak$ by $1-\betak$ in the proof of \rthe{mma}, one can also show that
\bear{mma9911c}
&&\pr (X(t+1)=1, X(t+2)=1, \ldots , X(t+M)=1) \nonumber\\
&&\ge ( r_1)^{M}=(\pr(X(1)=1))^M.
\eear
Letting $M=2$ in \req{mma9911c} yields the following corollary.

\bcor{posdep}
Suppose that the arriving process $\{X(t), t \ge 1\}$ satisfying (A1)-(A3).
Then $X(t+1)$ and $X(t+2)$ are positively correlated, i.e.,
\beq{mma9911d}
\ex[ X(t+1) X(t+2)] -\ex[X(t+1)]\ex[X(t+2)] \ge 0,
\eeq
where $\ex[X]$ denotes the expectation operator of the random variable $X$.
\ecor

There are two key properties used in the proof of \rthe{mma}: the regenerative property and Jensen's inequality (for convex functions).
To prove \rthe{geo}, we need the following generalization of Jensen's inequality.


\blem{hlp}
For any positive integers $t_1, t_2, \ldots, t_L$,
\beq{couple5555}
\prod_{\ell=1}^L \Big (\sum_{k=1}^K \pi_k (r_{0,k})^{t_\ell} \Big) \le \sum_{k=1}^K \pi_k (r_{0,k})^{\sum_{\ell=1}^L t_\ell}.
\eeq
\elem

Note that for $t_1=t_2=\cdots = t_L=1$, the inequality in \req{couple5555} reduces to Jensen's inequality for the convex function $x^{L}$ used in the proof of \rthe{mma}.

\bproof
Consider a  random variable $Y$ with the probability mass function $\pr (Y=r_{0,k})=\pi_k$, $k=1,2, \ldots, K$.
Since $r_{0,k} \ge 0$ for all $k$, $Y$ is nonnegative.
Then the right-hand-side of \req{couple5555} can be written as $\ex[Y^{\sum_{\ell=1}^L t_\ell}]$. Similarly, the left-hand-side of
\req{couple5555} can be written as $\prod_{\ell=1}^L \ex[Y^{t_\ell}]$. Thus, it suffices to show that
\beq{couple5555b}
\prod_{\ell=1}^L \Big ( \ex[Y^{t_\ell}] \Big) \le \ex[Y^{\sum_{\ell=1}^L t_\ell}].
\eeq
We show \req{couple5555b} by induction on $L$. For $L=2$, we consider two independent random variables $Y_1$ and $Y_2$ that have the same distribution as $Y$. Since $Y_1$ and $Y_2$ are nonnegative, for any two positive integers $t_1$ and $t_2$,
\beq{hlp1111}
(Y_1^{t_1}-Y_2^{t_1})(Y_1^{t_2}-Y_2^{t_2}) \ge 0.
\eeq
To see this, note that if $Y_1 \ge Y_2$, then $Y_1^{t_1} \ge Y_2^{t_1}$ and $Y_1^{t_2} \ge Y_2^{t_2}$.
Taking expectations on both side of \req{hlp1111} yields
\bear{hlp2222}
&&\ex[(Y_1^{t_1}-Y_2^{t_1})(Y_1^{t_2}-Y_2^{t_2})]\nonumber\\
&&=\ex[Y_1^{t_1+t_2}]-\ex[Y_2^{t_1}Y_1^{t_2}]-\ex[Y_2^{t_2}Y_1^{t_1}]+\ex[Y_2^{t_1+t_2}]\nonumber\\
&&\ge 0
\eear
Since $Y_1$ and $Y_2$ are independent and have the same distribution as $Y$,
we have from \req{hlp2222} that
\beq{hlp3333}\ex[Y^{t_1}] \ex[Y^{t_2}]\le \ex[ Y^{t_1+t_2}] .
\eeq
Now assume that \req{couple5555b} hold for $L-1$ as the induction hypothesis.
From \req{hlp3333} and the induction hypothesis, it follows that
\bear{hlp4444}
&&\ex[Y^{\sum_{\ell=1}^L t_\ell}] \nonumber\\
&&\ge \ex[Y^{\sum_{\ell=1}^{L-1} t_\ell}] \ex[Y^{t_L}] \nonumber\\
&&\ge \prod_{\ell=1}^L \Big ( \ex[Y^{t_\ell}] \Big).
\eear
\eproof

Now we prove \rthe{geo}.

\begin{figure}
	\centering
	\includegraphics[width=0.45\textwidth]{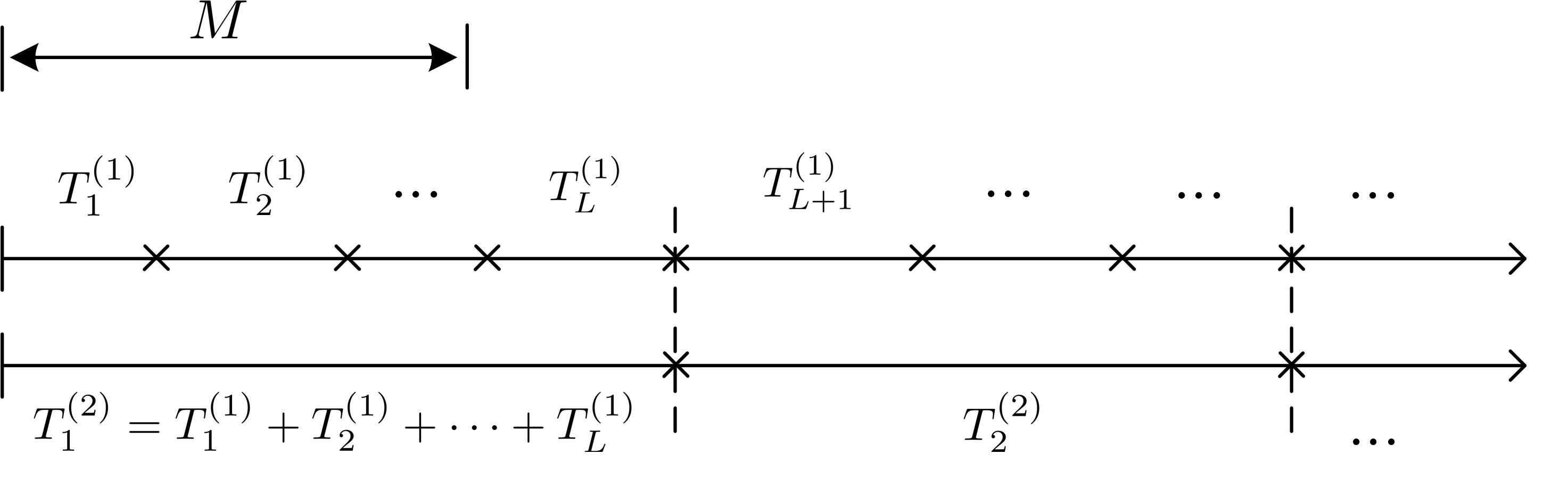}
	\caption{An illustration of coupling two sequences of group sizes $\{T^{(1)}_i, i \ge 1\}$ and $\{T^{(2)}_i, i \ge 1\}$.}
	\label{fig:coupling}
\end{figure}

\bproof(\rthe{geo})
To show that the expected relative cost in \req{dorf4444} is decreasing in $\omega$, it is equivalent to showing that
$\pr (X(1)=0, X(2)=0, \ldots , X(M)=0)$ is increasing in $\omega$. Consider two arrival processes
$\{X^{(1)}(t), t \ge 1\}$ and $\{X^{(2)}(t), t \ge 1\}$ that are generated by using the parameters $\omega_1$ and $\omega_2$ in (A$1^+$), respectively. Assume that $\omega_2 \ge \omega_1$.
Let $T^{(1)}_i$ (resp. $T^{(2)}_i$)  be the group size of the $i^{th}$ group in the first (resp. second) arrival process.
Note from (A$1^+$) that for all $i \ge 1$ and $n \ge 1$,
\bearn
P(T^{(1)}_i =n)&=&\omega_1^{n-1}(1-\omega_1), \\
P(T^{(2)}_i =n)&=&\omega_2^{n-1}(1-\omega_2).
\eearn

The trick of the proof is to couple the two sequences of group sizes $\{T^{(1)}_i, i \ge 1\}$ and $\{T^{(2)}_i, i \ge 1\}$ so that
the regenerative points of $\{X^{(2)}(t), t \ge 1\}$ is a subset of
the regenerative points of $\{X^{(1)}(t), t \ge 1\}$. Such a coupling is feasible because the random splitting of a renewal process with geometrically distributed interarrival times is also a renewal process
with geometrically distributed interarrival times.
In particular, the size of the first group for the second arrival process, i.e., $T^{(2)}_1$, is a sum of the sizes of several groups for the first arrival process, i.e.,
\beq{couple1111}
T^{(2)}_1 =\sum_{\ell=1}^L T^{(1)}_\ell,
\eeq
for some $L \ge 1$.
An illustration of coupling two sequences of group sizes $\{T^{(1)}_i, i \ge 1\}$ and $\{T^{(2)}_i, i \ge 1\}$ is shown in \rfig{coupling}.

Following the regenerative analysis in the proof of \rthe{mma}, we condition on the event $\{T^{(2)}_1=s\}$ and use the law of the total probability to derive that
\bear{couple9944}
&&\pr (X^{(2)}(1)=0,  \ldots , X^{(2)}(M)=0) \nonumber\\
&&=\sum_{s=1}^\infty \pr (X^{(2)}(1)=0,  \ldots , X^{(2)}(M)=0 |T^{(2)}_1=s)\nonumber\\
&&\quad\quad \pr (T^{(2)}_1=s).
\eear
For $s \ge M$, we have from (A3) that
\bear{cople3333}
&&\pr (X^{(2)}(1)=0,  \ldots , X^{(2)}(M)=0 |T^{(2)}_1=s)\nonumber\\
&&=\sum_{k=1}^K \pi_k (r_{0,k})^M.
\eear
From the coupling of these two arrival processes,
\bear{couple4444}
&&\pr (X^{(1)}(1)=0,  \ldots , X^{(1)}(M)=0 |T^{(2)}_1=s)\nonumber\\
&&=\pr (X^{(1)}(1)=0,  \ldots , X^{(1)}(M)=0 |\sum_{\ell=1}^L T^{(1)}_\ell=s)\nonumber\\
&&=\ex \Big [ \prod_{\ell=1}^L \Big (\sum_{k=1}^K \pi_k (r_{0,k})^{T^{(1)}_\ell} \Big)\big ].
\eear
As a direct consequence of \rlem{hlp}, we then have
\bear{couple6666}
&&\pr (X^{(2)}(1)=0,  \ldots , X^{(2)}(M)=0 |T^{(2)}_1=s)\nonumber\\
&&\ge \pr (X^{(1)}(1)=0,  \ldots , X^{(1)}(M)=0 |T^{(2)}_1=s).
\eear
The case for $s <M$ is similar, and we have from
\req{couple9944} that
\bear{couple7777}
&&\pr (X^{(2)}(1)=0,  \ldots , X^{(2)}(M)=0) \nonumber\\
&&\ge \pr (X^{(1)}(1)=0,  \ldots , X^{(1)}(M)=0).
\eear
\eproof

\bsubsec{Markov modulated processes}{mma}

In this section, we prove \rthe{mma2} and \rthe{mma4} by using the Markov property of Markov modulated processes (see, e.g., Chapter 8 and Chapter 9 of the book \cite{nelson2013probability}).

Recall that $Z(t)$ is the group type of the $t^{th}$ sample. In view of the memoryless property of the geometrical distribution, we know that with probability $\omega$, the $(t+1)^{th}$ sample is still in the same group of the $t^{th}$ sample.  With probability $1-\omega$, it is in another group.
Under (A$1^+$) and (A2),  the sequence of group types $\{Z(t), t =1,2, \ldots \}$ is a Markov chain with $K$ states.
Denote by $p_{i,j}$ the transition probability from state $i$ to state $j$ for the (hidden) Markov chain.
For such a Markov chain, we then have
\bear{trans1111}
p_{i,j}&=&\pr (Z(t+1)=j |Z(t)=i)\nonumber\\
&=&\left\{\begin{array}{ll}
	\omega+(1-\omega) \pi_i &\mbox{if $j=i$}\\
	(1-\omega) \pi_j &\mbox{if $j\ne i$}
\end{array}
\right..
\eear
It is easy to see that the correlation coefficient of $Z(t+1)$ and $Z(t)$ is simply $\omega$, i.e.,
\beq{mma1100}
\omega=\frac{\ex [Z(t+1) Z(t)]- \ex[Z(t+1)] \ex[Z(t)]}{\mbox{Var}(Z(t+1))\mbox{Var}(Z(t))}.
\eeq
From  (A3), we also know that  $\{X(t), t \ge 1\}$
is a Markov modulated process that is modualted by the (hidden) Markov chain $\{Z(t), t \ge 1\}$.
The conditional probability that $X(t)$ is {\em negative} given the (hidden) Markov chain is in the state $k$ is $\betak$, i.e.,
\beq{trans2222}
\betak=\pr (X(t)=0 |Z(t)=k) .
\eeq

As such, we have from the law of total probability that
\bear{trans3333}
&&\pr (X(1)=0, X(2)=0, \ldots , X(M)=0)\nonumber\\
&&= \sum_{k=1}^K\pr (X(1)=0,\ldots , X(M)=0 |Z(1)=k)\nonumber\\
&&\quad\quad\quad\pr (Z(1)=k).
\eear
From the (conditional) independence of Bernoulli samples in (A3), it follows that
\bear{trans3344}
&&\pr (X(1)=0, X(2)=0, \ldots , X(M)=0 |Z(1)=k)\nonumber\\
&&=\pr (X(2)=0, \ldots , X(M)=0 |Z(1)=k)\nonumber\\
&&\quad\quad \pr (X(1)=0|Z(1)=k) \nonumber\\
&&=\pr (X(2)=0, \ldots , X(M)=0 |Z(1)=k)\betak.
\eear
Using \req{trans3344} in  \req{trans3333} yields
\bear{trans3366}
&&\pr (X(1)=0, X(2)=0, \ldots , X(M)=0) \nonumber\\
&&=\sum_{k=1}^K\pr (X(2)=0, \ldots , X(M)=0 |Z(1)=k) \betak \pi_k. \nonumber\\
\eear

Now let
\beq{trans4444}
s_{k,M-1}=\pr (X(2)=0, \ldots , X(M)=0 |Z(1)=k).
\eeq
Similar to the argument for \req{trans3366}, we can further condition on the event $\{Z(2)=j\}$ and use the law of total probability to show that
\beq{trans5555}
s_{k,M-1}= \sum_{j=1}^K s_{j,M-2}r_{0,j} p_{k,j},
\eeq
for $k=1,2, \ldots, K$.
Let ${\bf s}_{M-1}$ be the $K \times 1$ (column) vector with its $k^{th}$ element being $s_{k,M-1}$,
${\bf P}=(p_{i,j})$ be the $K \times K$ transition probability matrix, and ${\bf R}$ be the diagonal matrix with
the $k^{th}$ diagonal element being $r_{0,k}$.
Then \req{trans5555} can be rewritten in the following matrix form:
\beq{trans6666}
{\bf s}_{M-1}={\bf P} {\bf R} {\bf s}_{M-2}.
\eeq
Since $s_{k,0}=1$ for all $k$, we have from \req{trans6666} that
\beq{trans7777}
{\bf s}_{M-1}
=  { ( {\bf P}{\bf R} )^{M-1} } {\bf 1},
\eeq
where ${\bf 1}$ is the $K \times 1$ vector with all its elements being 1.

Let ${\bf \pi}$ be the $1 \times K$ (row) vector with its $k^{th}$ element being $\pi_k$.
Then we have from \req{trans3366} and \req{trans7777} that
\bear{trans7777b}
&&\pr (X(1)=0, X(2)=0, \ldots , X(M)=0) \nonumber\\
&&= {\bf \pi} {\bf R}  { ( {\bf P}{\bf R} )^{M-1} } {\bf 1}.
\eear

Thus,
the expected relative cost is
\beq{trans8888}
\frac{M+1}{M}-{\bf \pi} {\bf R}  { ( {\bf P}{\bf R} )^{M-1} } {\bf 1},
\eeq
as in \rthe{mma2}.

For $\omega=1$, we note that the Markov chain $\{Z(t), t=1,2,\ldots \}$ stays at the same state from time 1 onward, and the $M$ random variables $\{X(1), X(2),\ldots X(M)\}$ are {\em i.i.d.} when conditioning on $Z(1)$. As such, they are exchangeable random variables, and the distribution of $\sum_{t=1}^M X(t)$ can be expressed as a mixture of Binomial distributions. For the special case $\omega=1$, our model of Markov modulated processes recovers the model of exchangeable binary random variables in \cite{lendle2012group} (see Assumptions 2 and 3 in \cite{lendle2012group}).

Now we prove \rthe{mma4}.
	
	\bproof (\rthe{mma4})
	Analogous to the proof of \rthe{mma}, it suffices to show that for any $M \ge 1$,
	\bear{mma9911u}
	&& \pr (X(1)=0, X(2)=0, \ldots , X(M)=0) \\ \nonumber
	&& \le r_0 (\omega+(1-\omega) r_0)^{M-1}.
	\eear
	
	For this, we  show that \req{mma9911u} holds 
	by induction on $M$. Since $\pr(X(1)=0)=r_0$,
	the inequality in \req{mma9911u} holds trivially for $M=1$.
	Assume that the inequality in
	\req{mma9911u} holds for all $s \le M-1$ as the induction hypothesis.
	From the law of total probability, we have
	\bear{mma9944u}
	&&\pr (X(1)=0,  \ldots , X(M)=0) \nonumber\\
	&=&\sum_{s=1}^\infty \pr (X(1)=0,  \ldots , X(M)=0 |T_1=s)\pr (T_1=s) \nonumber\\
	&=&\sum_{s=1}^{M-1} \pr (X(1)=0,  \ldots , X(M)=0 |T_1=s)\pr (T_1=s) \nonumber\\
	&+&\sum_{s=M}^\infty \pr (X(1)=0,  \ldots , X(M)=0 |T_1=s)\pr (T_1=s).\nonumber\\
	\eear
	Conditioning on the event $\{T_1=s\}$ for $s \ge M$, the number of samples in the first group is not smaller  than $M$.
	Thus, for $s\ge M$, we have from (A2) and (A3) that
	\bear{mma9955u}
	&&\pr (X(1)=0,  \ldots , X(M)=0 |T_1=s)\nonumber\\
	&&=\sum_{k=1}^K \pi_k (\betak)^M  \nonumber\\
	&&\le \sum_{k=1}^K \pi_k \betak =r_0,
	\eear
	where the last inequality follows from the fact that the convex function $x^M \le x$ for $0 \le x \le 1$.
	For $T_1=s \le M-1$, we know that the second group starts from $s+1$.
	It then follows from the renewal property in (A1) that
	\bear{mma9966u}
	&&\pr (X(1)=0,  \ldots , X(M)=0 |T_1=s)\nonumber\\
	&&=\pr (X(1)=0,  \ldots , X(s)=0 |T_1=s)\nonumber\\
	&&\quad\pr (X(s+1)=0,  \ldots , X(M)=0 |T_1=s)\nonumber\\
	&&=\Big (\sum_{k=1}^K \pi_k (\betak)^s \Big ) \pr (X(1)=0,  \ldots , X(M-s)=0 ) \nonumber\\
	&&\le \Big(\sum_{k=1}^K \pi_k \betak\Big)\pr (X(1)=0,  \ldots , X(M-s)=0 ) \nonumber\\
	&&\le  (r_0) \Big (r_0 (\omega+(1-\omega) r_0)^{M-s-1} \Big),
	\eear
	where the second last inequality follows from the fact that the convex function $x^M \le x$ for $0 \le x \le 1$,
	and the last inequality follows from
	the induction hypothesis.
	Since $T_1$ is geometrically distributed from (A$1^+$), we 
	have
	\beq{mma9915u}
	\pr (T_1=s)=(1-\omega)\omega^{s-1}.
	\eeq
	Using \req{mma9955u}, \req{mma9966u} and \req{mma9915u}
	in \req{mma9944u} yeilds
	\bear{mma9917u}
	&&\pr (X(1)=0,  \ldots , X(M)=0) \nonumber\\
	&\le &\sum_{s=1}^{M-1} (r_0) (r_0 (\omega+(1-\omega) r_0)^{M-s-1}) (1-\omega)\omega^{s-1} \nonumber\\
	&+&\sum_{s=M}^\infty r_0 (1-\omega)\omega^{s-1}\nonumber\\
	&=&r_0^2 (1-\omega) (\omega+(1-\omega) r_0)^{M-2} \nonumber \\
	&& \sum_{s=1}^{M-1}\Big (\frac{\omega}{\omega+(1-\omega) r_0}\Big)^{s-1} \nonumber \\
	&+&r_0 \omega^{M-1} \nonumber\\
	&=&r_0 (\omega+(1-\omega) r_0)^{M-1}
	\eear
	This then
	completes the induction in \req{mma9911u}.
	\eproof

\bsec{Pooled Testing with a Social Graph}{graph}


In the previous section, we consider samples in a line of a testing site, where the correlations between two contiguous samples are characterized by a line graph.
In this section, we extend the dependency between two samples to a general graph. Suppose that there is a social network  modeled by a graph $G=(V,E)$, where $V$ is the set of nodes, and $E$ is the set of edges. A node in $G$ represents a person in the social graph, and an edge between two persons represents frequent social contacts between these two persons. As a contagious disease can propagate the disease from an infected person to another person through the social contacts between these two persons, two persons connected by an edge are likely to infect each other. Thus, two samples obtained from two persons connected by an edge are also likely to be positively correlated.

The question for pooled testing with a social graph $G=(V,E)$ is how to exploit positive correlation from the edge connections in a social graph to save pooled testing costs. Intuitively, a  set of nodes that are densely connected to each other are likely to be positively correlated. In social network analysis (see, e.g., \cite{newman2010networks}), such a set of nodes is called a {\em community}.
In view of this, our idea for addressing the pooled testing problem with a social graph is to detect communities in a graph and then pool samples in the same community together for pooled testing.

Like pooled testing for people in a line, we define a {\em pooling strategy} for a graph $G=(V,E)$ with $n$ nodes, i.e., $|V|=n$, as a {\em permutation} $\sigma$ of $\{1,2, \ldots, n\}$  that puts the $n$ nodes into a line. As such, when we use the Dorfman two-stage algorithm with a given group size $M$, we can pool nodes $\sigma(1), \ldots, \sigma(M)$ in the first group, nodes $\sigma(M+1), \ldots, \sigma(2M)$ in the second group, etc.
A random pooling strategy for a graph $G=(V,E)$ is the strategy where the permutation $\sigma$ is selected at random among the $n!$ permutations.
The main objective of this section is to propose a pooling strategy from a community detection algorithm in \cite{chang2011general,chang2015relative,chang2017probabilistic} that can achieve a
lower expected relative cost than the random pooling strategy.

\bsubsec{The probabilistic framework of sampled graphs}{resampled}

In this section, we briefly review the probabilistic framework of sampled graphs for structural analysis in
\cite{chang2011general,chang2015relative,chang2017probabilistic}.
For a graph $G(V,E)$ with $n$ nodes, we index the $n$ nodes from $1,2,\ldots, n$. Also, let $A=(a_{i,j})$ be the $n \times n$ adjacency matrix of the graph, i.e.,
\bearn
a_{i,j}
=  \left\{\begin{array}{ll}
	1, & \mbox{if there is an edge from  node {\em i} to node {\em j}}, \\
	0, & $otherwise$.
\end{array} \right.
\eearn

Let $R_{u,w}$ be the set of paths from $u$ to $w$ and $R=\cup_{u,w \in V} R_{u,w}$ be the set of paths in the graph $G(V,E)$.  According to a probability mass function $p(\cdot)$, called the {\em path sampling distribution},
a path $r \in R$ is selected at random with probability $p(r)$.
Let $U$ (resp. $W$) be the starting (resp. ending) node of a randomly selected path by using the path sampling distribution $p(\cdot)$.
Then the bivariate distribution
\beq{frame1111}
p_{U,W}(u,w)=\pr(U=u,W=w)=\sum_{r\in R_{u,w}} p(r)
\eeq
is the probability that the ordered pair of two nodes $(u,w)$ is selected.
Intuitively, one might  interpret the bivariate distribution $p_{U,W}(u,w)$  in \req{frame1111} as the probability that both nodes $u$ and $w$ are infected (through one of the paths $r$ in $R_{u,w}$).
Thus, the bivariate distribution $p_{U,W}(u,w)$ can also be viewed as a similarity measure from node $u$ to node $w$  and this leads to the definition of a sampled graph in \cite{chang2011general,chang2015relative,chang2017probabilistic}.

\bdefin{sampled}{\bf (Sampled graph \cite{chang2011general,chang2015relative,chang2017probabilistic})} A  graph $G(V,E)$ that is sampled by randomly selecting an ordered pair of two nodes $(U,W)$ according to a specific bivariate distribution $p_{U,W}(\cdot,\cdot)$  in \req{frame1111} is called a {\em sampled graph} and it is denoted by the two-tuple $\sgo$.
\edefin

\begin{definition}\label{def:covariance}{\bf (Covariance and Community \cite{chang2015relative,chang2017probabilistic}))}
	For a sampled graph $\sgo$, the covariance between two nodes $u$ and $w$ is defined as follows:
	\begin{equation}\label{eq:exp8888}
	\qq (u,w)=p_{U,W} (u,w)-p_{U}(u) p_{W}(w).
	\end{equation}
	Moreover, the covariance between two sets $S_1$ and $S_2$ is defined as follows:
	\begin{equation}\label{eq:exp9999}
	\qq (S_1,S_2)=\sum_{u \in S_1}\sum_{w \in S_2}\qq (u,w).
	\end{equation}
	Two sets $S_1$ and $S_2$ are said to be positively correlated if $\qq (S_1,S_2)\ge 0$. In particular, if a subset of nodes $S \subset V$ is positively correlated to itself, i.e., $\qq(S,S)\ge 0$, then it is called a {\em community}.
\end{definition}

There are many  methods to obtain a sampled graph \cite{chang2015relative}.
In this paper, we will use the following bivariate distribution
\beq{link2234}
p_{U,W}(u,w)=c \cdot (A+0.5*A^2)(u,w),
\eeq
where $A=(a_{i,j})$ is the adjacency matrix of a graph $G=(V,E)$, and $c$ is the normalization constant so that the sum of $p_{U,W}(u,w)$ over $u$ and $w$ equals to 1. As such bivariate distribution is obtained from sampling paths with lengths 1 and 2, it seems to be a good sampling distribution for modelling the disease propagation within the second neighbors of an infected person.

\bsubsec{The hierarchical agglomerative algorithm for pooled testing in a graph}{community}

We propose a pooling strategy that uses the hierarchical agglomerative algorithm for community detection in sampled graphs
\cite{chang2017probabilistic}. The detailed steps are outlined in Algorithm \ref{alg:hierarchicalgroup}.
Initially, every node in the input graph is assigned to a set (community) that contains the node itself.
Then the algorithm recursively merges two sets that have the largest covariance into a new set. This is done by appending one set to the end of the other set so that the order of the elements in each set can be preserved.
Each merge of two sets reduces the number of sets by 1. Eventually, there is only one remaining set, and the order of the elements in the remaining set is the pooling strategy from the algorithm. It was shown in \cite{chang2017probabilistic} that all the sets are indeed {\em communities} if Algorithm \ref{alg:hierarchicalgroup} stops at the point when there does not exist a pair of two positively correlated sets. However, as our objective is to output a permutation for a pooling strategy, we continue the merge of two sets until there is only one remaining set.

\begin{algorithm}[t]
	\KwIn{A sampled graph $\sgo$.}
	
	\KwOut{A pooling strategy $\sigma$.}
	
	\noindent {\bf (H1)}  Initially, the number of sets $C$ is set to be $n$, and node $i$ is assigned to  the $i^{th}$ set, i.e., $S_i=\{i\}$, $i=1,2,\ldots, n$.
	
	\noindent {\bf (H2)} Compute the covariance $\qq(S_i, S_j)=\qq(\{i\},\{j\})$ from \req{exp8888} for all $i, j=1,2, \ldots, n$.
	
	\While{$C>1$}{
		\noindent {\bf  (H3)} Find the pairs of two sets $i$ and $j$ that have the largest covariance $\qq(S_i, S_j)$.
		
		\noindent {\bf  (H4)} Merge $S_i$ and $S_j$ into a new set $S_k$ by appending $S_j$ to $S_i$.
		
		\noindent {\bf  (H5)} Update the covariances as follows:
		\beq{random4400}\qq(S_k, S_k)= \qq(S_i, S_i)+2 \qq (S_i, S_j) + \qq (S_j, S_j).\eeq
		\For{each $\ell \ne k$}{
			\beq{random4455}\qq (S_k, S_\ell) =\qq(S_\ell, S_k)=\qq(S_i, S_\ell)+\qq(S_j, S_\ell).\eeq
		}
		$C=C-1$.
	}
	\noindent {\bf  (H6)} There is only one remaining set. Output $\sigma$ by letting $\sigma(i)$ be the $i^{th}$ element in the remaining set.
	\caption{The Hierarchical Agglomerative Algorithm for Pooled Testing with a Social Graph}
	\label{alg:hierarchicalgroup}
\end{algorithm}

\begin{figure}
	\centering
	\includegraphics[width=0.45\textwidth]{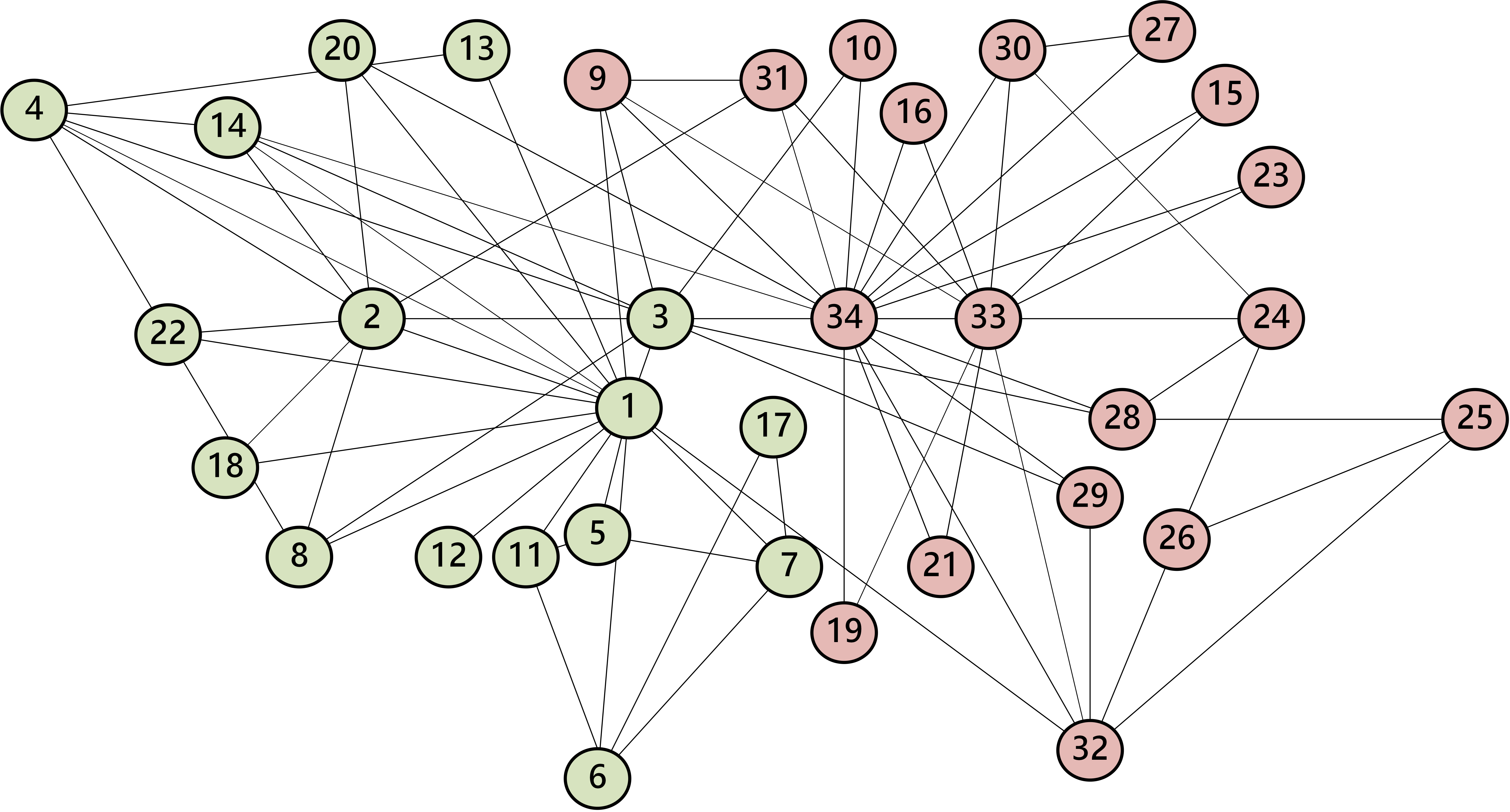}
	\caption{The Zachary karate club friendship network.}
	\label{fig:karate}
\end{figure}

\begin{figure} 
	\centering
	\subfloat[The dendrogram from Algorithm \ref{alg:hierarchicalgroup}.\label{fig:dendro}]{%
		\includegraphics[width=0.99\linewidth]{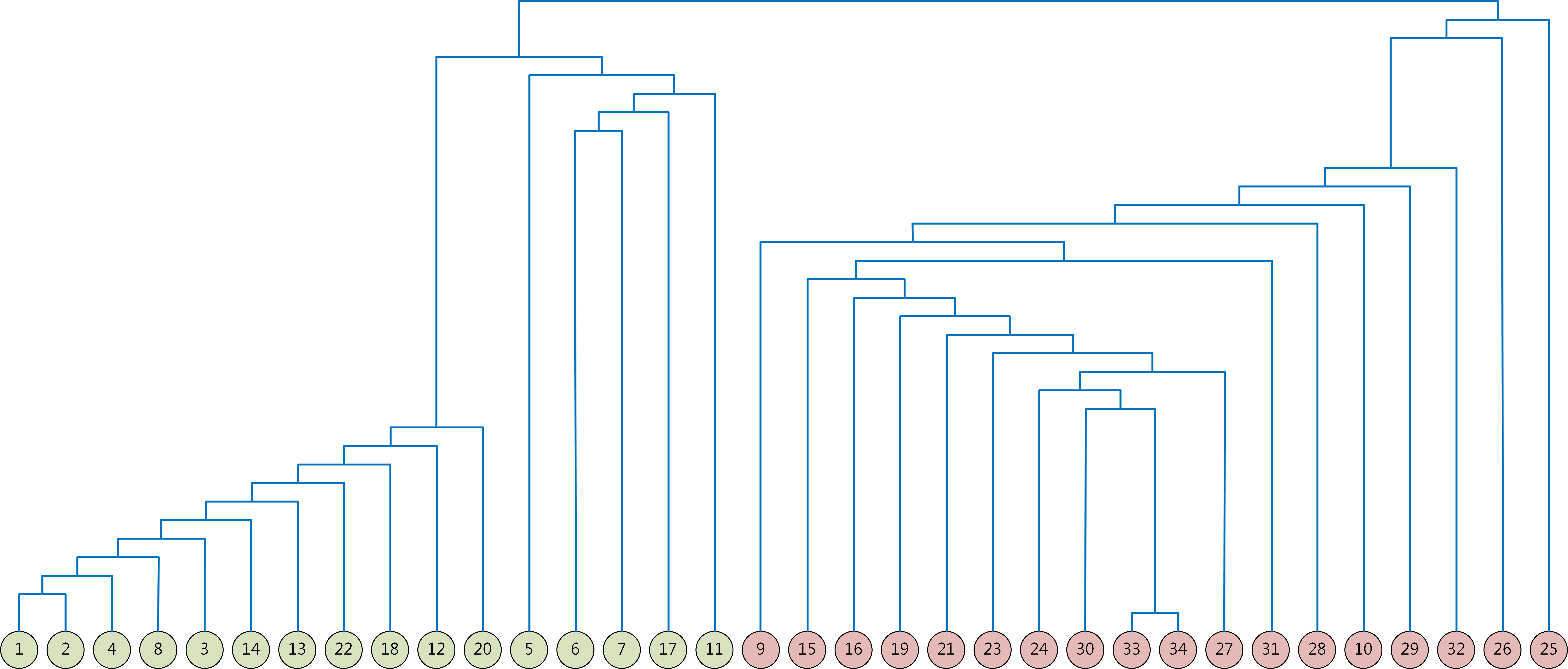}}
	\\
	\subfloat[Convert the dendrogram to a queue.\label{fig:den2que}]{%
		\includegraphics[width=0.99\linewidth]{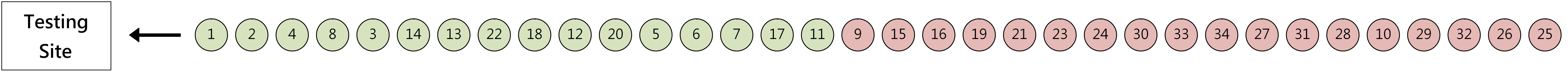}}
	\caption{(a) The dendrogram from Algorithm \ref{alg:hierarchicalgroup} for the Zachary karate club friendship network by using the similarity measure in \req{link2234}. (b) An illustration of the 34 members of the Zachary karate club forming a line to be tested in a testing site.}
	\label{fig:den_que} 
\end{figure}

As an illustrating example of our algorithm, we use the Zachary karate club friendship network \cite{zachary1977information}.
Such a friendship network is obtained by Wayne Zachary over the course of two years
in the early 1970s at an American university (see \rfig{karate}). During the course of the study, the club split into
two clusters (marked with two different colors in \rfig{karate}) because of a dispute between its administrator
(node 34) and its instructor (node 1).
In \rfig{dendro}, we show the dendrogram obtained from Algorithm \ref{alg:hierarchicalgroup} for the Zachary karate club friendship network by using the similarity measure in \req{link2234}.
A dendrogram for a hierarchical agglomerative algorithm is a tree-like graph with the height indicating the order of the merges of two sets.
The pooling strategy is the list of the 34 nodes in the bottom of this figure.
In \rfig{den2que}, we illustrate the members of the Zachary karate club forming a line to be tested in a testing site.

\bsec{Numerical Results}{num}

\bsubsec{Pooled testing on a line of a testing site}{site}

In this section, we compare the expected relative cost of Dorfman's two-stage method with that of a sophisticated group testing method in \cite{lin2020comparisons} by considering the special case with $K=2$, $r_{0,1}=0$ and $r_{0,2}=1$.
In this case, there are two types of arriving groups, and such a group is of type 1 (resp. type 2) with probability $\pi_1=r_1$ (resp. $\pi_2=r_0$).
The sizes of these arriving groups are i.i.d. geometric random variables with parameter $1-\omega$.
Moreover, with probability 1, samples in the type 1 group are positive and those in the type 2 group are negative.
Consequently, we have $X(t)=2-Z(t)$ for all $t$ and it reduces to the serial correlated model in \cite{hung1999robustness}.
The expected relative cost in this case is
\beq{cor2222}
\frac{M+1}{M}-r_0 (\omega+(1-\omega) r_0)^{M-1},
\eeq
where $r_0 = \pi_2$.
Notice that from \rthe{mma4}, \req{cor2222} achieves the lower bound of the expected relative costs under (A$1^+$), (A2), and (A3).

The optimal group size of $M$ that induces the lowest expected relative cost in \req{cor2222} can be determined by the prevalence rate $r_1$ and the parameter $\omega$ in the hidden Markov model. 
In general, the parameter $\omega$ is unknown and difficult to estimate; thus, in \rsubsubsec{nonoptimal}, we choose the group size $M$ according to that in Table I of \cite{dorfman1943detection}, which only depends on the prevalence rate $r_1$. 
However, if one can estimate the parameter $\omega$ reliably, the optimal group size of $M$ can be selected accordingly to further reduce the expected relative cost.
We optimize $M$ depending on both $r_1$ and $\omega$ in \rsubsubsec{optimal}.

\bsubsubsec{Group size $M$ determined by $r_1$}{nonoptimal}
In this section, we choose the group size $M$ from Table I of \cite{dorfman1943detection} that only depends on the prevalence rate $r_1$ (since the parameter $\omega$ in the hidden Markov model is generally unknown).		

We numerically evaluate the expected relative cost in \req{cor2222} for each value of $r_1$ ranging from $1\%$ to $10\%$ with increment of $1\%$, and each value of $\omega$ ranging from $0$ to $0.9$ with increment of $0.1$.
The results are shown in \rtab{cost}.
To compare the expected relative costs of Dorfman's two-stage algorithm (with positively correlated samples) with those of the $(d_1,d_2)$-regular pooling matrices \cite{lin2020comparisons},
we also list the lowest expected relative costs of the $(d_1,d_2)$-regular pooling matrices (Table I of \cite{lin2020comparisons}) in \rtab{cost}.
In this table, we can easily verify that the expected relative cost decreases in $\omega$.
The numbers given in boldface are the expected relative costs of Dorfman's two-stage algorithm of the smallest values of $\omega$ that outperform those of the $(d_1,d_2)$-regular pooling matrices under the same prevalence rate $r_1$.
We can observe that when the prevalence rate $r_1$ is low (e.g., $r_1<5\%$), the gain by Dorfman two-stage method is not as good as that of $(d_1,d_2)$-regular pooling matrix, except for some large $\omega$.
The reason is that under a low prevalence rate, there are very few positive samples in a group, and such positive samples can be detected easily by using the sophisticated group testing method, thus saving more testing costs. However, Dorfman's 2-stage algorithm can only check if the group contains at least one positive sample at the first stage. When a group of $M$ samples includes any positive ones (even if there is only one positive sample in the group), all the $M$ samples should be retested individually at the second stage.
Thus, the performance of Dorfman's method is not as good as those of sophisticated group testing methods, on the premise that the prevalence rate is low and correlations between samples in a group are small.
But when the prevalence rate $r_1$ is high (e.g., $r_1 \geq 5\%$), the simple Dorfman's method can achieve better performance with some moderate positive correlation $\omega$.

To show the advantage of using positively correlated samples in Dorfman's two-stage method, we calculate the ratio of the expected relative cost with the positive correlation $\omega$ to that of the i.i.d. Bernoulli samples ($\omega =0$) in \rtab{saving}.
For example, under the prevalence rate $r_1=1\%$, the expected relative cost with $\omega=0.1$ is $0.1865$ from \rtab{cost}, and thus the ratio is $0.1865/0.1956=95.4\%$.

\begin{table*}[h]
	\centering
	\caption{The expected relative cost of the Dorfman two-stage algorithm with group size $M$ and the lowest expected relative cost of $(d_1,d_2)$-regular in \cite{lin2020comparisons}. The numbers given in boldface are the expected relative costs of Dorfman's two-stage algorithm of the smallest values of $\omega$ that outperform those of the $(d_1,d_2)$-regular pooling matrices under the same prevalence rate $r_1$.}
	\label{tab:cost}
	\resizebox{\textwidth}{!}{
		\begin{tabular}{|c|c|c|c|c|c|c|c|c|c|c|c|c|c|c|}
			\hline
			& \multicolumn{11}{c|}{The Dorfman Two-stage Algorithm} & $(d_1,d_2)$-regular \\ \hline
			$r_1$ & $M$ & $\omega=0$ & $\omega=0.1$ & $\omega=0.2$ & $\omega=0.3$ & $\omega=0.4$ & $\omega=0.5$ & $\omega=0.6$ & $\omega=0.7$ & $\omega=0.8$ & $\omega=0.9$ & Lowest Cost \\ \hline
			1\% & 11 & 0.1956 & 0.1865 & 0.1773 & 0.1681 & 0.1587 & 0.1493 & 0.1398 & 0.1302 & \textbf{0.1205} & 0.1108 & 0.1218 \\ \hline
			2\% & 8 & 0.2742 & 0.2620 & 0.2496 & 0.2371 & 0.2244 & 0.2116 & 0.1986 & \textbf{0.1854} & 0.1721 & 0.1586 & 0.1881  \\ \hline
			3\% & 6 & 0.3337 & 0.3207 & 0.3076 & 0.2943 & 0.2809 & 0.2673 & \textbf{0.2535} & 0.2395 & 0.2254 & 0.2111 & 0.2545  \\ \hline
			4\% & 6 & 0.3839 & 0.3675 & 0.3507 & 0.3337 & 0.3165 & \textbf{0.2989} & 0.2810 & 0.2629 & 0.2445 & 0.2257 & 0.3147  \\ \hline
			5\% & 5 & 0.4262 & 0.4098 & 0.3931 & 0.3762 & \textbf{0.3590} & 0.3415 & 0.3238 & 0.3057 & 0.2874 & 0.2689 & 0.3678  \\ \hline
			6\% & 5 & 0.4661 & 0.4472 & 0.4279 & \textbf{0.4082} & 0.3882 & 0.3678 & 0.3470 & 0.3259 & 0.3043 & 0.2824 & 0.4166  \\ \hline
			7\% & 5 & 0.5043 & 0.4831 & \textbf{0.4615} & 0.4393 & 0.4167 & 0.3935 & 0.3699 & 0.3457 & 0.3210 & 0.2958 & 0.4627  \\ \hline
			8\% & 4 & 0.5336 & 0.5148 & \textbf{0.4956} & 0.4761 & 0.4562 & 0.4360 & 0.4155 & 0.3947 & 0.3735 & 0.3519 & 0.5035  \\ \hline
			9\% & 4 & 0.5643 & 0.5437 & \textbf{0.5227} & 0.5014 & 0.4796 & 0.4574 & 0.4348 & 0.4117 & 0.3883 & 0.3643 & 0.5416  \\ \hline
			10\% & 4 & 0.5939 & \textbf{0.5718} & 0.5492 & 0.5261 & 0.5025 & 0.4784 & 0.4537 & 0.4286 & 0.4029 & 0.3767 & 0.5760 \\ \hline
		\end{tabular}
	}
\end{table*}

\begin{table}[h]
	\centering
	\caption{The ratio of the expected relative cost with positive correlation $\omega$ to that of the i.i.d. Bernoulli samples ($\omega=0$) under different prevalence rate $r_1$. (unit: \%)}
	\label{tab:saving}
	\scalebox{0.9}{
		\begin{tabular}{|c|c|c|c|c|c|c|c|c|c|}
			\hline
			$r_1\backslash \omega$ & $0.1$ & $0.2$ & $0.3$ & $0.4$ & $0.5$ & $0.6$ & $0.7$ & $0.8$ & $0.9$ \\ \hline
			1\% & 95.4 & 90.7 & 85.9 & 81.2 & 76.3 & 71.5 & 66.6 & 61.6 & 56.6 \\ \hline
			2\% & 95.5 & 91.0 & 86.5 & 81.8 & 77.2 & 72.4 & 67.6 & 62.8 & 57.8 \\ \hline
			3\% & 96.1 & 92.2 & 88.2 & 84.2 & 80.1 & 76.0 & 71.8 & 67.6 & 63.3 \\ \hline
			4\% & 95.7 & 91.4 & 86.9 & 82.4 & 77.9 & 73.2 & 68.5 & 63.7 & 58.8 \\ \hline
			5\% & 96.1 & 92.2 & 88.3 & 84.2 & 80.1 & 76.0 & 71.7 & 67.4 & 63.1 \\ \hline
			6\% & 95.9 & 91.8 & 87.6 & 83.3 & 78.9 & 74.5 & 69.9 & 65.3 & 60.6 \\ \hline
			7\% & 95.8 & 91.5 & 87.1 & 82.6 & 78.0 & 73.3 & 68.5 & 63.7 & 58.6 \\ \hline
			8\% & 96.5 & 92.9 & 89.2 & 85.5 & 81.7 & 77.9 & 74.0 & 70.0 & 65.9 \\ \hline
			9\% & 96.4 & 92.6 & 88.9 & 85.0 & 81.1 & 77.1 & 73.0 & 68.8 & 64.6 \\ \hline
			10\% & 96.3 & 92.5 & 88.6 & 84.6 & 80.5 & 76.4 & 72.2 & 67.8 & 63.4 \\ \hline
		\end{tabular}
	}
\end{table}

\bsubsubsec{Group size $M$ determined by $r_1$ and $\omega$}{optimal}
	
	In this section, the optimal group size $M$ that induces the lowest expected relative cost is determined by both the prevalence rate $r_1$ and the correlation coefficient $\omega$.
	For each value of $r_1$ ranging from $1\%$ to $10\%$ with increment of $1\%$, and each value of $\omega$ ranging from $0$ to $0.9$ with increment of $0.1$,
	we show its optimal group size $M$ in \rtab{size_opt} and its corresponding expected relative cost in \rtab{cost_opt}.
	Intuitively, with correlated samples, the group size for pooled testing can be larger. This can be verified in \rtab{size_opt}, which shows the size $M$ increases in $\omega$ for a fixed value of $r_1$.
	To make a comparison of the expected relative costs of Dorfman's two-stage algorithm (with positively correlated samples) and those of the $(d_1,d_2)$-regular pooling matrices \cite{lin2020comparisons},
	we also list the lowest expected relative costs of the $(d_1,d_2)$-regular pooling matrices (Table I of \cite{lin2020comparisons}) in \rtab{cost_opt}.
	To show the advantage of using positively correlated samples in Dorfman's two-stage method, we calculate the ratio of the expected relative cost with the positive correlation $\omega$ to that of the i.i.d. Bernoulli samples ($\omega =0$) in \rtab{saving_opt}.

\begin{table}[h]
	\centering
	\caption {The optimal group size of the Dorfman two-stage algorithm with different values of the prevalence rate $r_1$ and the correlation coefficient $\omega$.}
	\label{tab:size_opt}
	\scalebox{0.95}{
		\begin{tabular}{|c|c|c|c|c|c|c|c|c|c|c|c|c|}
			\hline
			$r_1 \backslash \omega$ & $0$ & $0.1$ & $0.2$ & $0.3$ & $0.4$ & $0.5$ & $0.6$ & $0.7$ & $0.8$ & $0.9$ \\ \hline
			1\% & 11 & 11 & 12 & 12 & 13 & 15 & 16 & 19 & 23 & 32 \\ \hline
			2\% & 8 & 8 & 8 & 9 & 10 & 11 & 12 & 14 & 16 & 23 \\ \hline
			3\% & 6 & 7 & 7 & 7 & 8 & 9 & 10 & 11 & 14 & 19 \\ \hline
			4\% & 6 & 6 & 6 & 7 & 7 & 8 & 9 & 10 & 12 & 17  \\ \hline
			5\% & 5 & 5 & 6 & 6 & 6 & 7 & 8 & 9 & 11 & 15  \\ \hline
			6\% & 5 & 5 & 5 & 6 & 6 & 6 & 7 & 8 & 10 & 14  \\ \hline
			7\% & 5 & 5 & 5 & 5 & 6 & 6 & 7 & 8 & 9 & 13 \\ \hline
			8\% & 4 & 4 & 5 & 5 & 5 & 6 & 6 & 7 & 9 & 12 \\ \hline
			9\% & 4 & 4 & 4 & 5 & 5 & 5 & 6 & 7 & 8 & 12 \\ \hline
			10\% & 4 & 4 & 4 & 4 & 5 & 5 & 6 & 7 & 8 & 11 \\ \hline
		\end{tabular}
	}
\end{table}

\begin{table*}[h]
	\centering
	\caption{The expected relative cost of the Dorfman two-stage algorithm with its optimal group size in \rtab{size_opt}, and the lowest expected relative cost of $(d_1,d_2)$-regular in \cite{lin2020comparisons}. The numbers given in boldface are the expected relative costs of Dorfman's two-stage algorithm of the smallest values of $\omega$ that outperform those of the $(d_1,d_2)$-regular pooling matrices under the same prevalence rate $r_1$.}
	\label{tab:cost_opt}
	\resizebox{\textwidth}{!}{
		\begin{tabular}{|c|c|c|c|c|c|c|c|c|c|c|c|}
			\hline
			& \multicolumn{10}{c|}{The Dorfman Two-stage Algorithm with Positively Correlated Samples} & $(d_1,d_2)$-regular \\ \hline
			$r_1$ & $\omega=0$ & $\omega=0.1$ & $\omega=0.2$ & $\omega=0.3$ & $\omega=0.4$ & $\omega=0.5$ & $\omega=0.6$ & $\omega=0.7$ & $\omega=0.8$ & $\omega=0.9$ & Lowest Cost \\ \hline
			1\% & 0.1956 & 0.1865 & 0.1771 & 0.1670 & 0.1559 & 0.1438 & 0.1303 & \textbf{0.1147} & 0.0961 & 0.0715 & 0.1218 \\ \hline
			2\% & 0.2742 & 0.2620 & 0.2496 & 0.2356 & 0.2209 & 0.2046 & \textbf{0.1862} & 0.1652 & 0.1397 & 0.1057 & 0.1881 \\ \hline
			3\% & 0.3337 & 0.3198 & 0.3044 & 0.2888 & 0.2708 & \textbf{0.2516} & 0.2299 & 0.2048 & 0.1744 & 0.1337 & 0.2545 \\ \hline
			4\% & 0.3839 & 0.3675 & 0.3507 & 0.3333 & \textbf{0.3131} & 0.2916 & 0.2673 & 0.2388 & 0.2045 & 0.1585 & 0.3147 \\ \hline
			5\% & 0.4262 & 0.4098 & 0.3921 & 0.3717 & \textbf{0.3509} & 0.3267 & 0.3003 & 0.2693 & 0.2317 & 0.1810 & 0.3678 \\ \hline
			6\% & 0.4661 & 0.4472 & 0.4279 & \textbf{0.4082} & 0.3841 & 0.3595 & 0.3304 & 0.2972 & 0.2568 & 0.2022 & 0.4166 \\ \hline
			7\% & 0.5043 & 0.4831 & \textbf{0.4615} & 0.4393 & 0.4162 & 0.3884 & 0.3586 & 0.3234 & 0.2803 & 0.2221 & 0.4627 \\ \hline
			8\% & 0.5336 & 0.5148 & \textbf{0.4939} & 0.4694 & 0.4443 & 0.4165 & 0.3847 & 0.3476 & 0.3025 & 0.2411 & 0.5035 \\ \hline
			9\% & 0.5643 & 0.5437 & \textbf{0.5227} & 0.4985 & 0.4712 & 0.4431 & 0.4091 & 0.3707 & 0.3237 & 0.2595 & 0.5416 \\ \hline
			10\% & 0.5939 & \textbf{0.5718} & 0.5492 & 0.5261 & 0.4973 & 0.4669 & 0.4328 & 0.3932 & 0.3437 & 0.277 & 0.5760 \\ \hline
		\end{tabular}
	}
\end{table*}

\begin{table}[]
	\centering
	\caption{With optimal group sizes in \rtab{size_opt}, the ratio of the expected relative cost with positive correlation $\omega$ to that of the i.i.d. Bernoulli samples ($\omega=0$) under different prevalence rate $r_1$. (unit: \%)}
	\label{tab:saving_opt}
	\scalebox{0.9}{
		\begin{tabular}{|c|c|c|c|c|c|c|c|c|c|}
			\hline
			$r_1\backslash \omega$ & $0.1$ & $0.2$ & $0.3$ & $0.4$ & $0.5$ & $0.6$ & $0.7$ & $0.8$ & $0.9$ \\ \hline
			1\% & 95.4 & 90.5 & 85.4 & 79.7 & 73.5 & 66.6 & 58.7 & 49.2 & 36.6 \\ \hline
			2\% & 95.5 & 91.0 & 85.9 & 80.6 & 74.6 & 67.9 & 60.2 & 50.9 & 38.5 \\ \hline
			3\% & 95.8 & 91.2 & 86.6 & 81.2 & 75.4 & 68.9 & 61.4 & 52.3 & 40.1 \\ \hline
			4\% & 95.7 & 91.4 & 86.8 & 81.5 & 76.0 & 69.6 & 62.2 & 53.3 & 41.3 \\ \hline
			5\% & 96.1 & 92.0 & 87.2 & 82.3 & 76.7 & 70.5 & 63.2 & 54.4 & 42.5 \\ \hline
			6\% & 95.9 & 91.8 & 87.6 & 82.4 & 77.1 & 70.9 & 63.8 & 55.1 & 43.4 \\ \hline
			7\% & 95.8 & 91.5 & 87.1 & 82.5 & 77.0 & 71.1 & 64.1 & 55.6 & 44.0 \\ \hline
			8\% & 96.5 & 92.6 & 88.0 & 83.3 & 78.1 & 72.1 & 65.1 & 56.7 & 45.2 \\ \hline
			9\% & 96.4 & 92.6 & 88.4 & 83.5 & 78.5 & 72.5 & 65.7 & 57.4 & 46.0 \\ \hline
			10\% & 96.3 & 92.5 & 88.6 & 83.7 & 78.6 & 72.9 & 66.2 & 57.9 & 46.6 \\ \hline
		\end{tabular}
	}
\end{table}

\bsubsec{Pooled testing with a social graph}{poolgraph}

In this section, we report our simulation results for pooled testing with a social graph.
For our experiments, we use a synthetic dataset and three real-world datasets.
The synthetic dataset is constructed by the small-world model in \cite{watts1998collective} as follows. First, we generate a ring with $1,000$ nodes, and each node has a degree of $30$ connected to its nearest neighbors. Then, for each edge, with probability $0.5$, we remove that edge and add a new one to two randomly selected nodes. By doing so, we obtain the synthetic dataset.
The three real-world datasets are: the email-Eu-core in \cite{yin2017local} \cite{leskovec2007graph}, the political blogs in \cite{adamic2005political} and the ego-Facebook in \cite{leskovec2012learning}. 
There are 986 nodes and 16,064 edges for the email-Eu-core network after removing multiple edges, self-loops, and nodes with degree 0.
For the political blogs, there are 1,224 nodes and 16,715 edges.
For the ego-Facebook dataset, we remove multiple edges, self-loops, and nodes that are not in the largest component of the network as in \cite{lu2020explainable}. By doing so, there are $2,851$ nodes and $62,318$ edges left in the network. 
The basic information of datasets is given in Table \ref{tab:dataInfo}.

\begin{table}[]
	\centering
	\caption{ Basic information of four datasets. Note that the political blogs dataset is not connected; the average path length and the diameter of the largest connected component in political blogs are reported.}
	\label{tab:dataInfo}
	\scalebox{0.8}{
		\begin{tabular}{|c|c|c|c|c|}
			\hline
			Dataset & small-world & email-Eu-core & political blogs  & ego-Facebook\\ \hline
			Number of nodes & 1,000 & 986 & 1,224   & 2,851\\ \hline
			Number of edges & 15,000 & 16,064 & 16,715  & 62,318 \\ \hline
			Average degree  & 30 &  32.5842 & 27.3121 & 43.7166 \\ \hline
			Average excess degree & 29.3581 & 73.6564 & 80.2587   & 98.0664\\ \hline
			\begin{tabular}[c]{@{}c@{}}Average clustering\\ coefficient\end{tabular}  & 0.1133 & 0.4071 & 0.3197  &  0.5914\\ \hline
			Average path length & 2.4414 & 2.5843 & 2.7467  & 4.1353\\ \hline
			Diameter & 3 &  7 & 8   & 14\\ \hline
			Density & 3.0030e-2 & 3.3080e-2 & 2.2332e-2   & 1.5339e-2\\ \hline
		\end{tabular}
	}
\end{table}

We also need a model for modelling disease propagation in a network.
A widely used model is the independent cascade (IC) model  (see, e.g., Kempe, Kleinberg, and Tardos in \cite{kempe2003maximizing}).
In the IC model, an {\em infected} node can transmit the disease to a neighboring {\em susceptible} node (through an edge) with a certain propagation probability $\phi$. An infected neighboring node can continue the propagation of the disease to its neighbors.
For our experiments, a set of seeded nodes $S$ are randomly selected in the IC model. Each neighbor of a seeded node is infected with probability $\phi$. These infected nodes are called the first-generation cascade of a seeded node and they can continue infecting their neighbors.
The  $D$-generation {\em cascade} from a seeded node is generated by collecting the set of infected nodes within the distance $D$ of the seeded node, and the $D$-generation cascade from the set $S$ is generated by taking the union of  the $D$-generation cascades of the seeded nodes in $S$.
In our experiments, we set $\phi=0.1$ and $D=2$.

\begin{figure}
	\centering
	\includegraphics[width=0.4\textwidth]{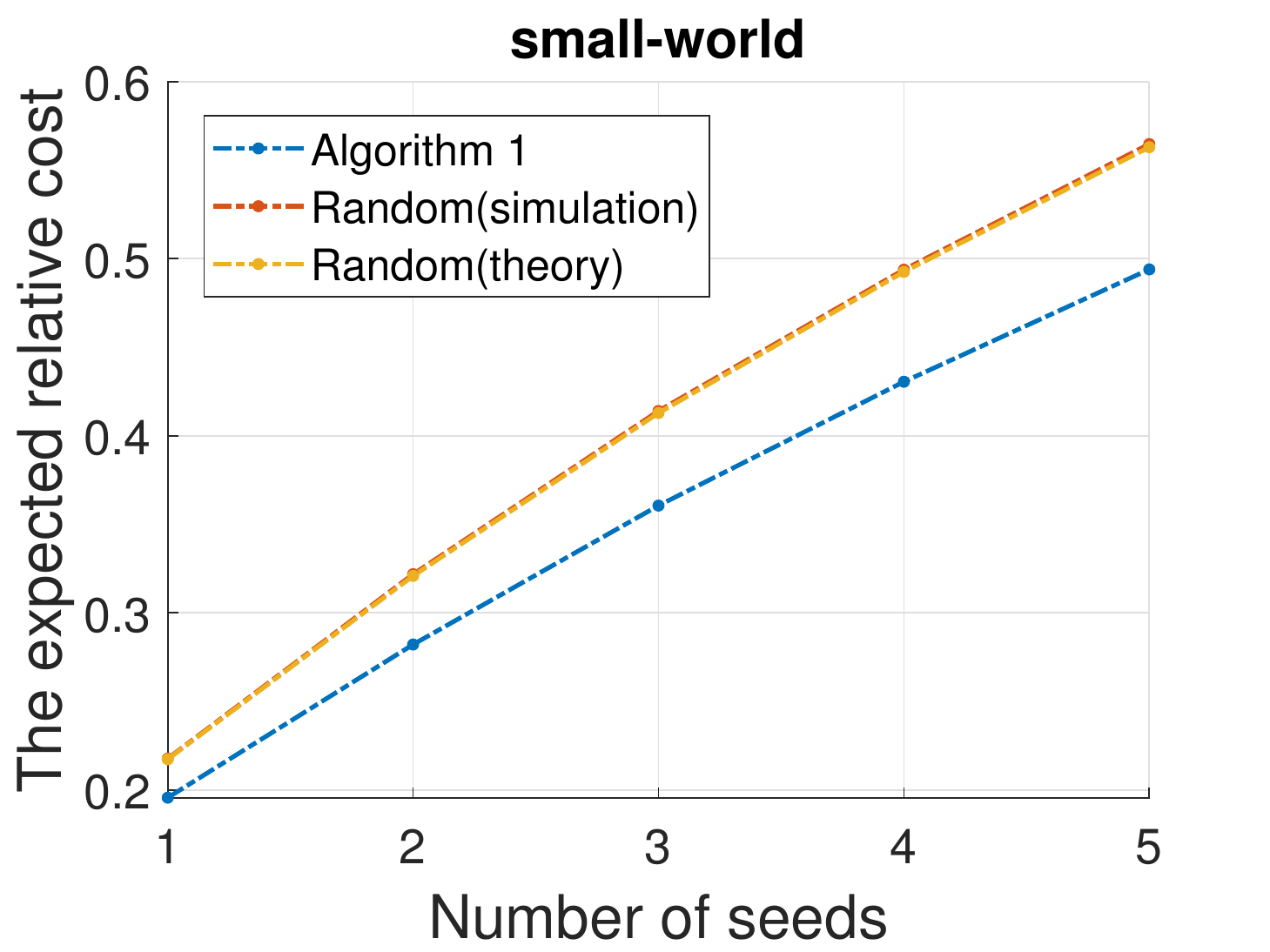}
	\caption{The expected relative cost of Dorfman's two-stage algorithm with $M=10$ as a function of the number of seeded nodes $|S|$ from 1 to 5 for the small-world dataset.}
	\label{fig:dorfsmall}
\end{figure}

\begin{figure}
	\centering
	\includegraphics[width=0.4\textwidth]{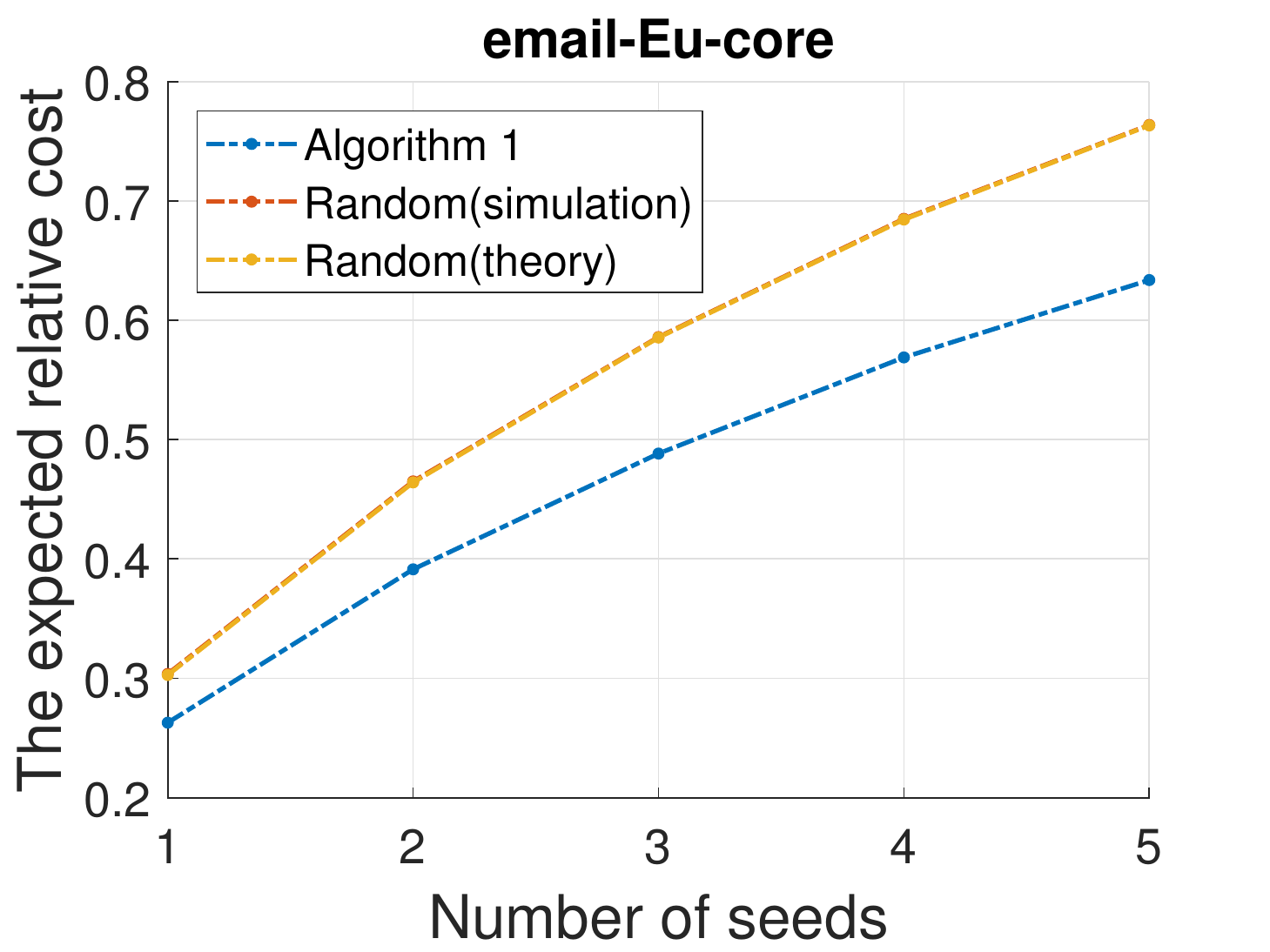}
	\caption{The expected relative cost of Dorfman's two-stage algorithm with $M=10$ as a function of the number of seeded nodes $|S|$ from 1 to 5 for the email-Eu-core dataset.}
	\label{fig:dorfemail}
\end{figure}

\begin{figure}
	\centering
	\includegraphics[width=0.4\textwidth]{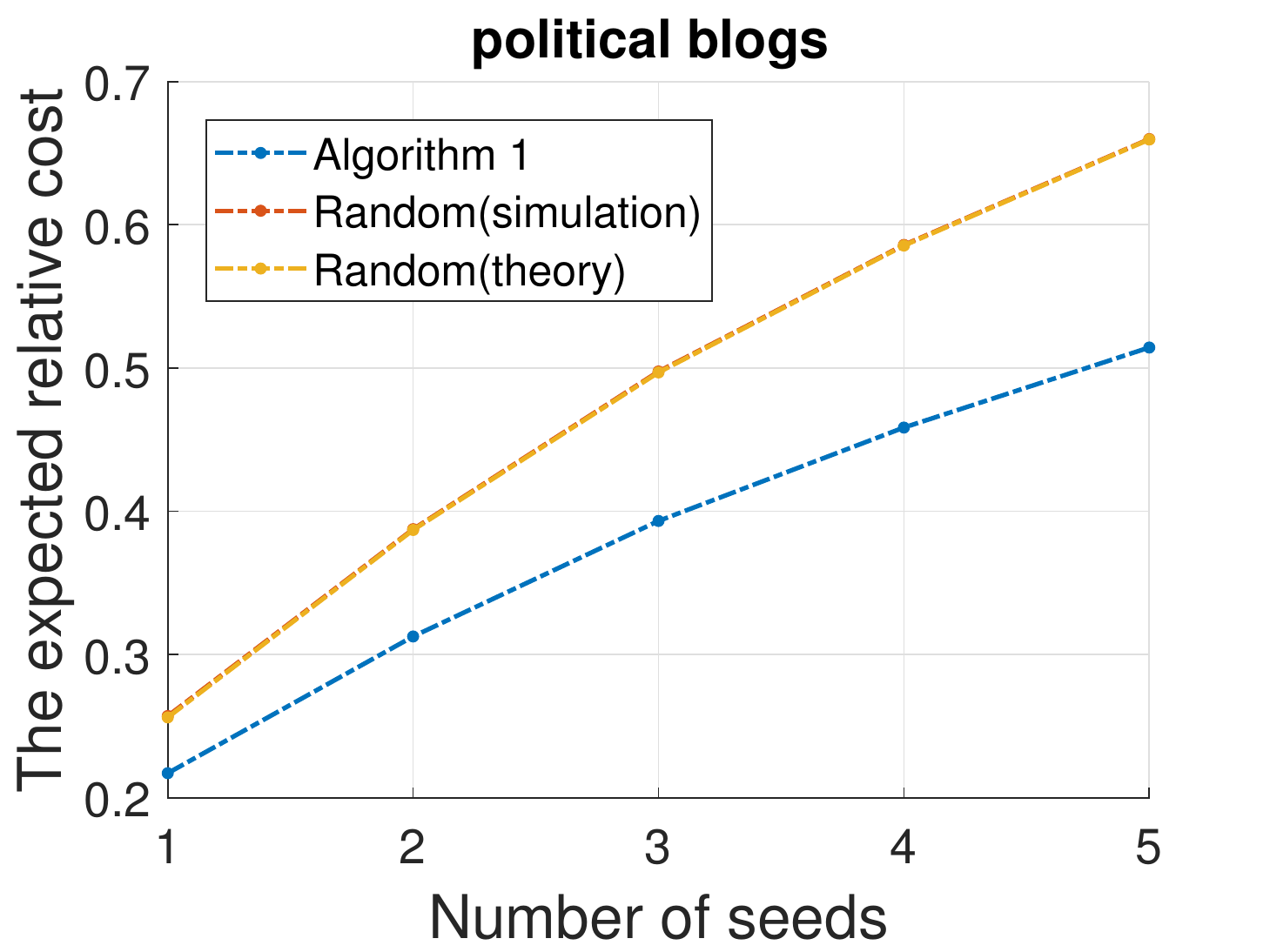}
	\caption{The expected relative cost of Dorfman's two-stage algorithm with $M=10$ as a function of the number of 	seeded nodes $|S|$ from 1 to 5 for the political blogs dataset.}
	\label{fig:dorfblogs}
\end{figure}

\begin{figure}
	\centering
	\includegraphics[width=0.4\textwidth]{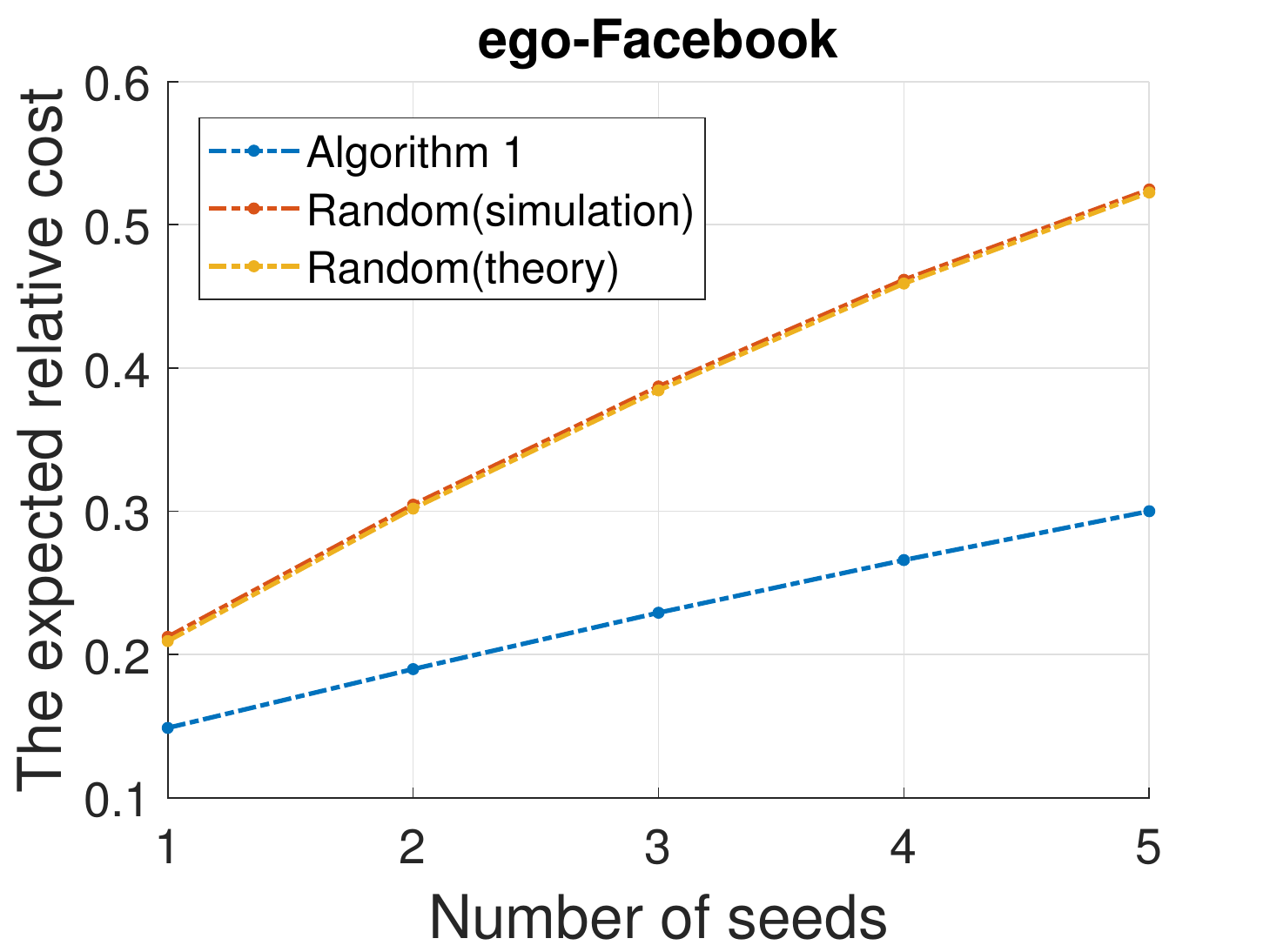}
	\caption{The expected relative cost of Dorfman's two-stage algorithm with $M=10$ as a function of the number of seeded nodes $|S|$ from 1 to 5 for the  ego-Facebook dataset.}
	\label{fig:dorffb}
\end{figure}

The pooling strategy for each dataset is obtained in the same way as that for the
Zachary karate club friendship network in \rsubsec{community}. Specifically, we first generate a sampled graph by using the bivariate distribution in \req{link2234}.
Then we use the hierarchical agglomerative algorithm for pooled testing with a social graph in Algorithm \ref{alg:hierarchicalgroup} to generate the pooling strategy.
In \rfig{dorfsmall} (resp. \rfig{dorfemail}, \rfig{dorfblogs}, \rfig{dorffb}), we show the expected relative cost of Dorfman's two-stage algorithm with the group size  $M=10$, as a function of the number of seeded nodes $|S|$ for the small-world dataset (resp. the email-Eu-core dataset, the political blogs dataset, the ego-Facebook dataset).
In our experiments, the number of seeded nodes $|S|$  is from 1 to 5.
Each data point is obtained from averaging 10,000 independent runs. Specifically, for the $i^{th}$ run, we measure the prevalence rate $r_1^{(i)}$  and the total number of tests $I^{(i)}$. The expected relative cost is calculated by $$\frac{\sum_{i=1}^{10,000}{I^{(i)}}}{n*10,000},$$
where $n$ is the number of nodes in the graph.
The average prevalence rate is calculated by
$$\frac{\sum_{i=1}^{10,000}r_1^{(i)}}{10,000}.$$

As shown in \rfig{dorfsmall}, the pooling strategy from Algorithm \ref{alg:hierarchicalgroup} results in much lower expected relative costs than those from the random pooling strategy.
We note that the two curves, Random(simulation) and Random(Theory) from \req{dorf1111}, are almost identical in this figure.
We confirm the same finding for the email-Eu-core, the political blogs and the ego-Facebook datasets in \rfig{dorfemail}, \rfig{dorfblogs} and \rfig{dorffb}.
To understand the effect of the number of seeded nodes in a dataset, we show the average prevalence rates in Table \ref{tab:dataPrev}.
As shown in this table, the prevalence rates are in the range of 1\% to 12\% that are basically in line with the prevalence rates of COVID-19 in various countries.
Moreover, we can observe that the email-Eu-core network has the highest prevalence rates among the four datasets.
Intuitively, the higher density and the higher averaging clustering coefficient, the higher the prevalence rate.
However, under the IC model, the total number of people infected in a network highly depends on the network's structure.
To conclude, under the IC model, the expected relative costs for the small-world dataset and the three real-world datasets can be significantly reduced by roughly 10\%-13\% and 20\%-35\%, respectively, by exploiting positive correlation within a social graph.


\begin{table}
	\centering
	\caption{Average prevalence rates (unit: \%).}
	\label{tab:dataPrev}
	\scalebox{0.95}{
		\begin{tabular}{|c|c|c|c|c|c|}
			\hline
			Dataset $\backslash$ Number of seeds & 1 & 2 & 3 & 4 & 5 \\ \hline
			small-world &  1.26 &  2.51 &  3.75 &  4.95 & 6.13 \\ \hline
			email-Eu-core & 2.63 &  5.15 &  7.42 & 9.62 &  11.68 \\ \hline
			political blogs & 1.91 & 3.77 & 5.56 & 7.21 & 8.78 \\ \hline
			ego-Facebook & 1.26 & 2.44 & 3.59 & 4.72 & 5.79 \\ \hline
		\end{tabular}
	}
\end{table}

\bsec{Conclusion}{con}

By modelling the arrival process of a COVID-19 testing site by a regenerative process, we showed that the expected relative cost for positively correlated samples is not higher than that of i.i.d. samples with the same prevalence rate. A more detailed model by a Markov modulated process allows us to derive a closed-form expression for the expected relative cost. Using the closed-form expression in \rthe{mma2}, we showed that for a specific Markov modulated process with a moderate positive correlation, the gain by Dorfman's two-stage method outperforms those by using sophisticated strategies with $(d_1,d_2)$-regular pooling matrices when the prevalence rate is higher than $5\%$.

One important extension of our results is to consider the pooled testing problem with a social graph. The frequent social contacts between two persons are connected by an edge in the social graph. To exploit positive correlation in a social graph, we adopted the probabilistic framework of sampled graphs for structural analysis in \cite{chang2011general,chang2015relative,chang2017probabilistic} and proposed a hierarchical agglomerative algorithm for pooled testing with a social graph in Algorithm \ref{alg:hierarchicalgroup}. Our numerical results show that the pooled testing strategy obtained from Algorithm \ref{alg:hierarchicalgroup} can have significant cost reduction (roughly 20\%-35\%) in comparison with random pooling when the Dorfman two-stage algorithm is used.

There are several possible extensions for our work:
\begin{description}
	\item[(i)] Association of random samples: in this paper, we model in the arrival process by three explicit assumptions. It is possible to further generalize our results by using the notion of association of random variables \cite{esary1967association}. In particular, it was shown in Theorem 4.1 of \cite{esary1967association} that \req{mma9911b} and \req{mma9911c} hold for associated binary random variables.
	\item[(ii)] Sensitivity/specificity analysis: in this paper, we did not consider the effect of noise. Noise (see, e.g., the monograph  \cite{aldridge2019group} for various noise models) can affect sensitivity (true positive rate) and specificity (true negative rate) of a testing method.   It would be of interest to see how the expected relative cost is affected by a certain type of noise, e.g., the dilution noise.
\end{description}

\begin{IEEEbiography}[{\includegraphics[width=1in,height=1.25in,clip,keepaspectratio]{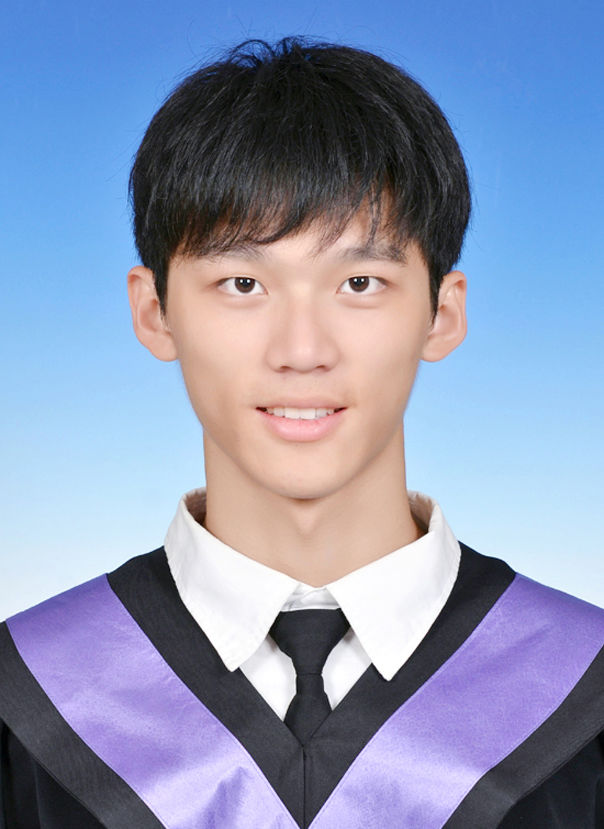}}]
	{Yi-Jheng Lin} received his B.S. degree in electrical engineering from National Tsing Hua University, Hsinchu, Taiwan, in 2018. He is currently pursuing the Ph.D. degree in the Institute of Communications Engineering, National Tsing Hua University, Hsinchu, Taiwan. His research interests include wireless communication and cognitive radio networks.
\end{IEEEbiography}

\begin{IEEEbiography}[{\includegraphics[width=1in,height=1.25in,clip,keepaspectratio]{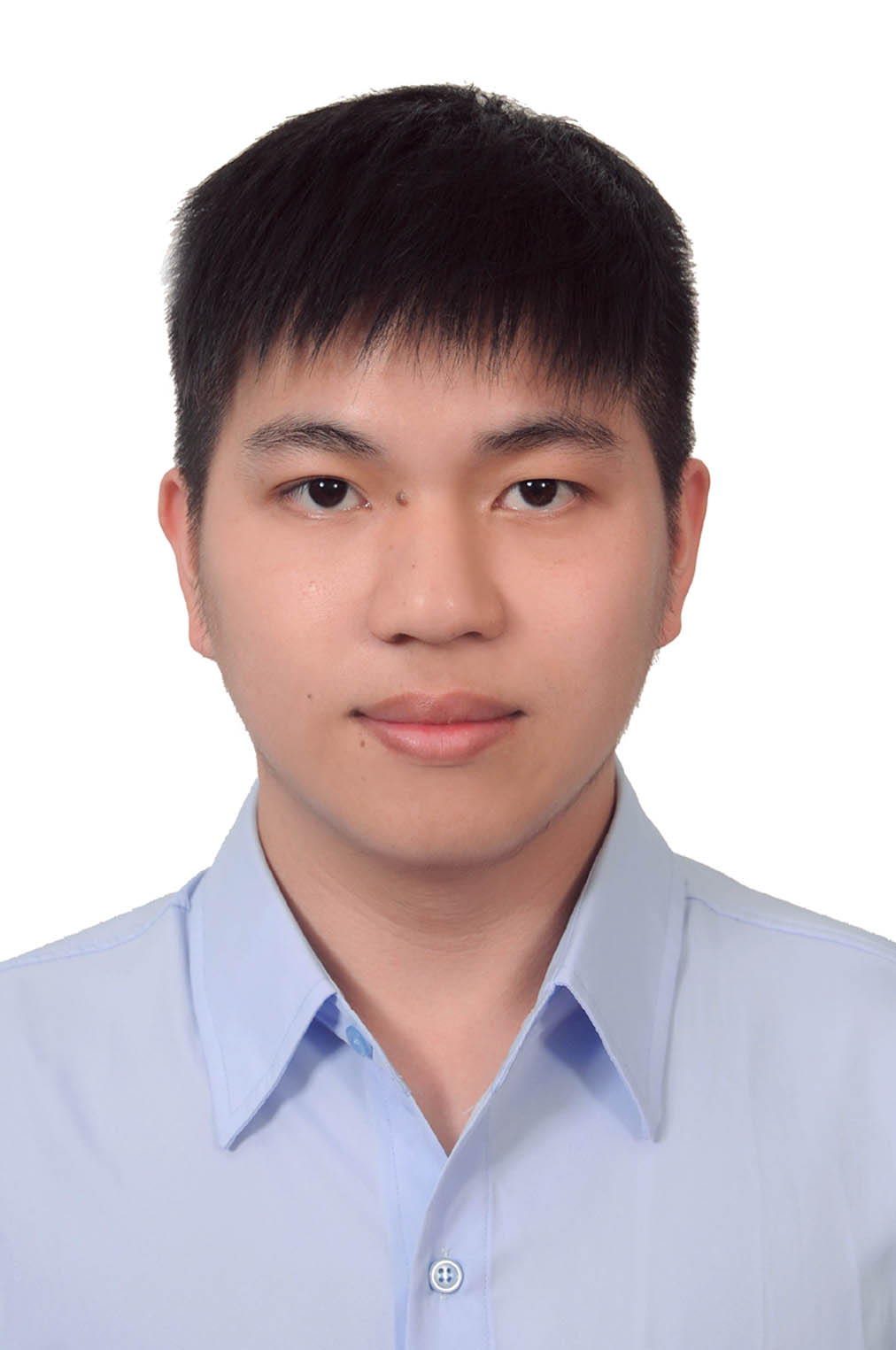}}]
{Che-Hao Yu} received his B.S. degree in mathematics from National Tsing-Hua University, Hsinchu, Taiwan (R.O.C.), in 2018, and the M.S. degree in communications engineering from National Tsing Hua University, Hsinchu, Taiwan (R.O.C.), in 2020. His research interest is in 5G wireless communication.
\end{IEEEbiography}

\begin{IEEEbiography}[{\includegraphics[width=1in,height=1.25in,clip,keepaspectratio]{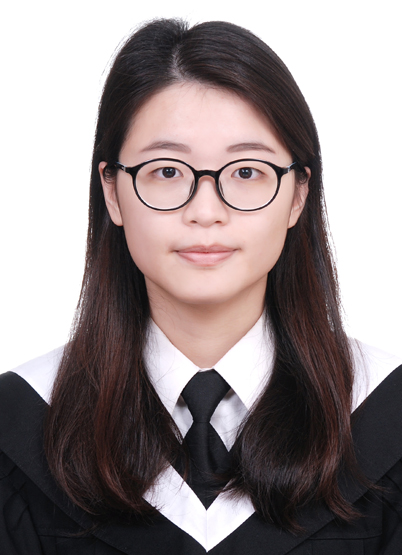}}]
	{Tzu-Hsuan Liu} received the B.S. degree in communication engineering from National Central University, Taoyuan, Taiwan (R.O.C.), in 2018. She is currently pursuing the M.S. degree in the Institute of Communications Engineering, National Tsing Hua University, Hsinchu, Taiwan (R.O.C.). Her research interest is in 5G wireless communication.
\end{IEEEbiography}

\begin{IEEEbiography}[{\includegraphics[width=1in,height=1.25in,clip,keepaspectratio]{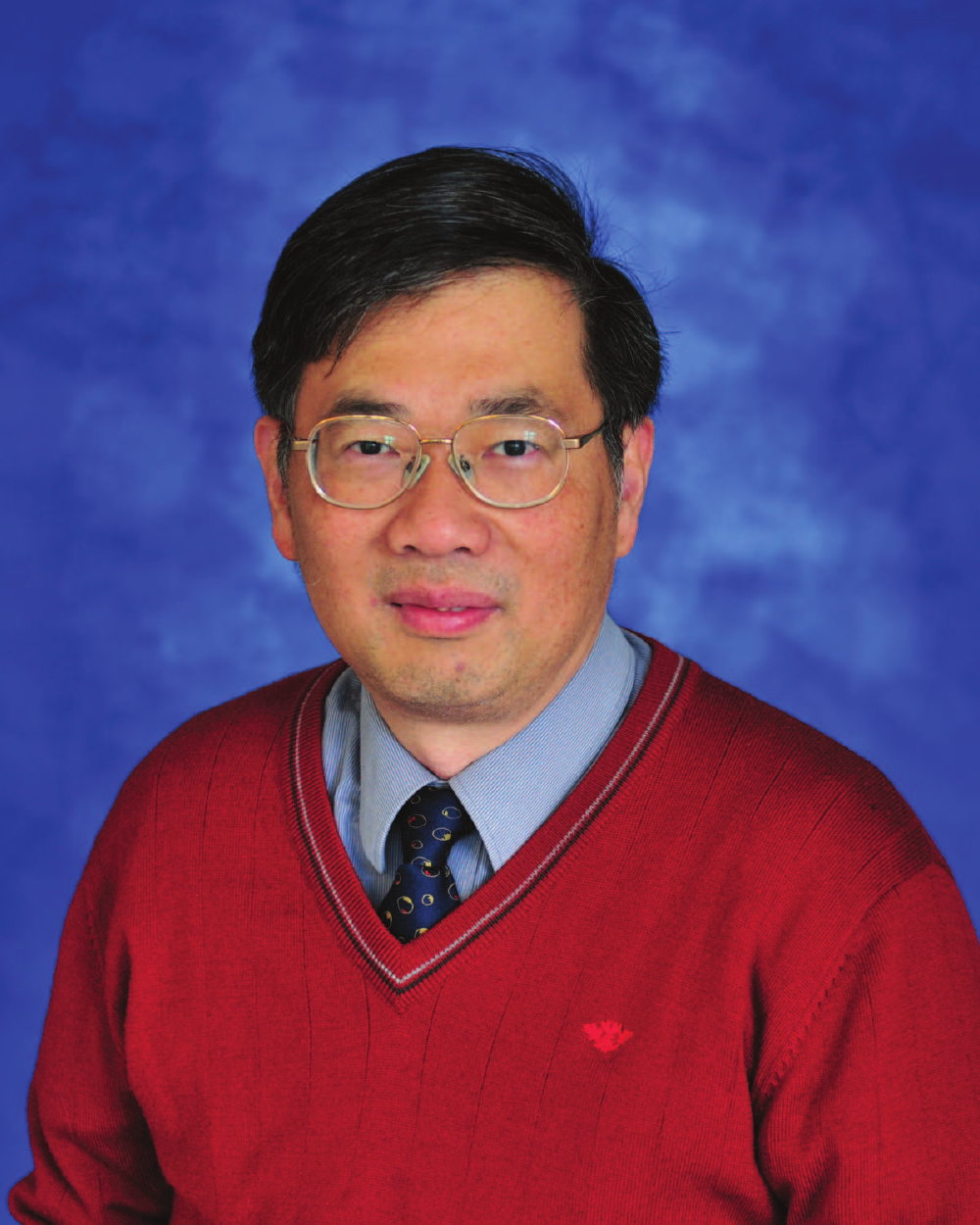}}]
	{Cheng-Shang Chang}
	(S'85-M'86-M'89-SM'93-F'04)
	received the B.S. degree from National Taiwan
	University, Taipei, Taiwan, in 1983, and the M.S.
	and Ph.D. degrees from Columbia University, New
	York, NY, USA, in 1986 and 1989, respectively, all
	in electrical engineering.
	
	From 1989 to 1993, he was employed as a
	Research Staff Member with the IBM Thomas J.
	Watson Research Center, Yorktown Heights, NY,
	USA. Since 1993, he has been with the Department
	of Electrical Engineering, National Tsing Hua
	University, Taiwan, where he is a Tsing Hua Distinguished Chair Professor. He is the author
	of the book Performance Guarantees in Communication Networks (Springer,
	2000) and the coauthor of the book Principles, Architectures and Mathematical
	Theory of High Performance Packet Switches (Ministry of Education, R.O.C.,
	2006). His current research interests are concerned with network science, big data analytics,
	mathematical modeling of the Internet, and high-speed switching.
	
	Dr. Chang served as an Editor for Operations Research from 1992 to 1999,
	an Editor for the {\em IEEE/ACM TRANSACTIONS ON NETWORKING} from 2007
	to 2009, and an Editor for the {\em IEEE TRANSACTIONS
		ON NETWORK SCIENCE AND ENGINEERING} from 2014 to 2017. He is currently serving as an Editor-at-Large for the {\em IEEE/ACM
		TRANSACTIONS ON NETWORKING}. He is a member of IFIP Working
	Group 7.3. He received an IBM Outstanding Innovation Award in 1992, an
	IBM Faculty Partnership Award in 2001, and Outstanding Research Awards
	from the National Science Council, Taiwan, in 1998, 2000, and 2002, respectively.
	He also received Outstanding Teaching Awards from both the College
	of EECS and the university itself in 2003. He was appointed as the first Y. Z.
	Hsu Scientific Chair Professor in 2002. He received the Merit NSC Research Fellow Award from the
	National Science Council, R.O.C. in 2011. He also received the Academic Award in 2011 and the National Chair Professorship in 2017 from
	the Ministry of Education, R.O.C. He is the recipient of the 2017 IEEE INFOCOM Achievement Award.
\end{IEEEbiography}

\begin{IEEEbiography}
	[{\includegraphics[width=1in,height=1.25in,clip,keepaspectratio]{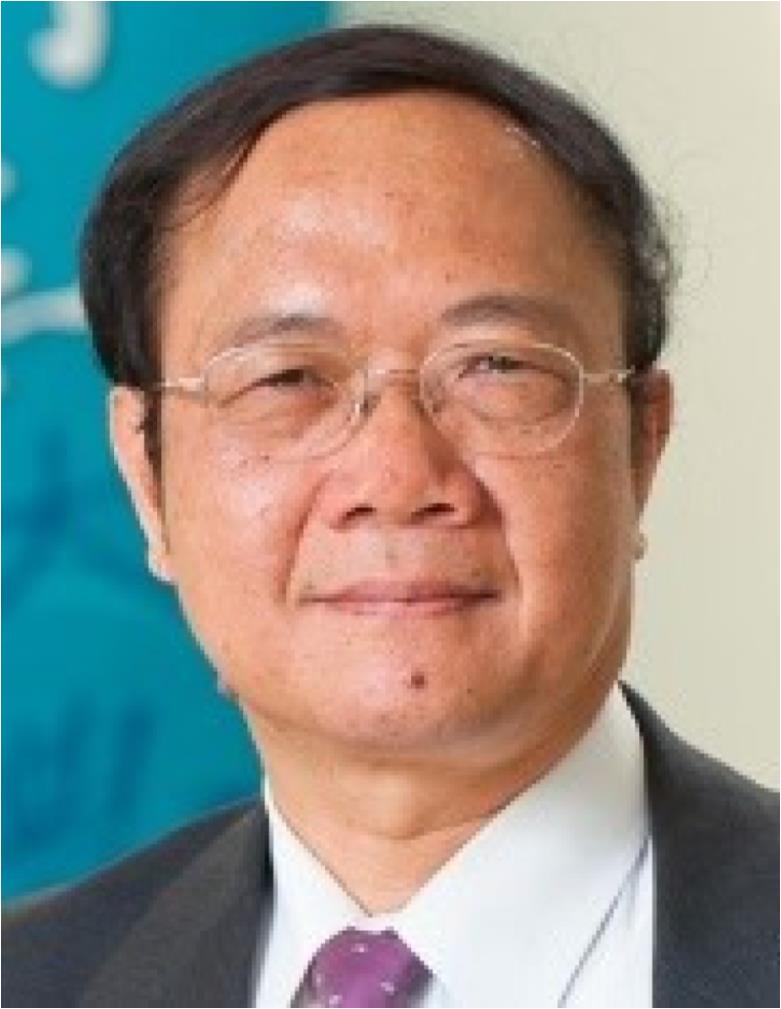}}]
{Wen-Tsuen Chen}(M'87-SM'90-F'94) received his
B.S. degree in nuclear engineering from National
Tsing Hua University, Taiwan, and M.S. and Ph.D.
degrees in electrical engineering and computer sciences both from University of California, Berkeley,
in 1970, 1973, and 1976, respectively. He has been
with the Department of Computer Science of National Tsing Hua University since 1976 and served
as Chairman of the Department, Dean of College of
Electrical Engineering and Computer Science, and
the President of National Tsing Hua University. In
March 2012, he joined the Academia Sinica, Taiwan as a Distinguished
Research Fellow of the Institute of Information Science until June 2018. Currently he is Sun Yun-suan Chair Professor of National Tsing Hua University.
His research interests include computer networks, wireless sensor networks,
mobile computing, and parallel computing. Dr. Chen received numerous
awards for his academic accomplishments in computer networking and parallel
processing, including Outstanding Research Award of the National Science
Council, Academic Award in Engineering from the Ministry of Education,
Technical Achievement Award and Taylor L. Booth Education Award of the
IEEE Computer Society, and is currently a lifelong National Chair of the
Ministry of Education, Taiwan. Dr. Chen is the Founding General Chair of
the IEEE International Conference on Parallel and Distributed Systems and the
General Chair of the IEEE International Conference on Distributed Computing
Systems. He is an IEEE Fellow and a Fellow of the Chinese Technology
Management Association.
\end{IEEEbiography}

\end{document}